%% file: main.tex
\documentclass[sigplan,10pt]{acmart}

\AtBeginDocument{%
  \providecommand\BibTeX{{%
    \normalfont B\kern-0.5em{\scshape i\kern-0.25em b}\kern-0.8em\TeX}}}

\copyrightyear{2024} 
\acmYear{2024} 
\setcopyright{acmlicensed}\acmConference[EuroSys '24]{Nineteenth European Conference on Computer Systems}{April 22--25, 2024}{Athens, Greece}
\acmBooktitle{Nineteenth European Conference on Computer Systems (EuroSys '24), April 22--25, 2024, Athens, Greece}
\acmPrice{15.00}
\acmDOI{10.1145/3627703.3629585}
\acmISBN{979-8-4007-0437-6/24/04}

\acmSubmissionID{\#452}

\usepackage{tabularx}

\usepackage{multirow}
\usepackage{makecell}
\usepackage{hhline}
\usepackage{xcolor}
\usepackage{bm}

\usepackage{amsmath}
\usepackage{graphicx}
\usepackage{enumitem}
\usepackage{caption}
\usepackage{subcaption}
\usepackage[ruled,linesnumbered,noend]{algorithm2e}

\SetCommentSty{mycommfont}
\usepackage{amsthm}
\usepackage{ifthen}
\usepackage{comment}

\usepackage{listings}
\definecolor{deepblue}{rgb}{0,0,0.5}
\definecolor{deepred}{rgb}{0.6,0,0}
\definecolor{deepgreen}{rgb}{0,0.5,0}
\graphicspath{ {./figures/} }

\newcommand{\SystemName}{DynaPipe}

\renewcommand{\paragraph}[1]{\vspace{1mm} \noindent \textbf{#1} \thickspace}

\definecolor{mtinstr}{HTML}{FF8000}
\definecolor{mtresponse}{HTML}{FF6666}

\begin{document}

\title{\SystemName: Optimizing Multi-task Training through Dynamic Pipelines
}

\author{Chenyu Jiang}
\email{jchenyu@connect.hku.hk}
\authornote{Work done during Chenyu’s internship at AWS.}
\orcid{0009-0006-5714-3872}
\affiliation{%
  \institution{The University of Hong Kong}
  \country{}
}

\author{Zhen Jia}
\email{zhej@amazon.com}
\affiliation{%
  \institution{Amazon Web Services}
  \country{}
}

\author{Shuai Zheng}
\email{shuai@boson.ai}
\authornote{Work done while at AWS.}
\affiliation{%
  \institution{Boson AI}
  \country{}
}

\author{Yida Wang}
\email{wangyida@amazon.com}
\affiliation{%
  \institution{Amazon Web Services}
  \country{}
}

\author{Chuan Wu}
\email{cwu@cs.hku.hk}
\affiliation{%
  \institution{The University of Hong Kong}
  \country{}
}


\begin{abstract}
Multi-task model training has been adopted to enable a single deep neural network model (often a large language model) to handle multiple tasks (e.g., question answering and text summarization).
Multi-task training commonly receives input sequences of highly different lengths due to the diverse contexts of different tasks.
Padding (to the same sequence length) or packing (short examples into long sequences of the same length) is usually adopted to prepare input samples for model training, which is nonetheless not space or computation efficient.
This paper proposes a dynamic micro-batching approach to tackle sequence length variation and enable efficient multi-task model training.
We advocate pipeline-parallel training of the large model with variable-length micro-batches, each of which potentially comprises a different number of samples.
We optimize micro-batch construction using a dynamic programming-based approach, and handle micro-batch execution time variation through dynamic pipeline and communication scheduling, enabling highly efficient pipeline training.
Extensive evaluation on the FLANv2 dataset demonstrates up to 4.39x higher training throughput when training T5, and 3.25x when training GPT, as compared with packing-based baselines. \SystemName{}'s source code is publicly available at \url{https://github.com/awslabs/optimizing-multitask-training-through-dynamic-pipelines}.
\end{abstract}




\maketitle

\input{sections/introduction}

\input{sections/background_motivation}

\input{sections/overview}

\input{sections/microbatch_partition}

\input{sections/schedule_optimization}

\input{sections/async_comm_planning}

\input{sections/other_optimizations}

\input{sections/evaluation}

\bibliographystyle{ACM-Reference-Format}
\bibliography{references}

\clearpage
\input{sections/artifact_appendix}

\end{document}

%% file: sections/introduction.tex
\section{Introduction}

Recent studies have shown that a single deep neural network (DNN), e.g., a large language model (LLM), can be trained/fine-tuned on a mixture of datasets to perform multiple tasks effectively~\cite{brown2020gpt3, mishra2022instrtune, wei2022flan, sanh2022t0}.
For example, T0~\cite{sanh2022t0} is fine-tuned on 62 different NLP datasets and can perform a wide-range of tasks including question answering, sentiment analysis, summarization and sentence completion.
Flan-T5 and Flan-PaLM~\cite{wei2022flan} are fine-tuned on 473 datasets from 146 categories of tasks.

\begin{figure}[t]
     \centering
     \begin{subfigure}[b]{\linewidth}
         \centering
         \includegraphics[width=\textwidth]{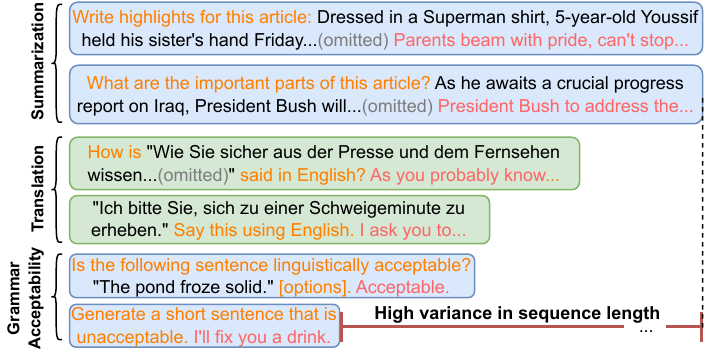}
         \caption{Example input sequences in multi-task training (instruction tuning~\cite{mishra2022instrtune}). \textcolor{mtinstr}{Orange} texts are instructions to the model. Inputs to process are colored \textbf{black}. Expected responses are in \textcolor{mtresponse}{red}.
         }
         \label{fig:example_variation}
     \end{subfigure}
     \hfill
     \begin{subfigure}[b]{\linewidth}
        \centering
        \includegraphics[width=0.9\linewidth]{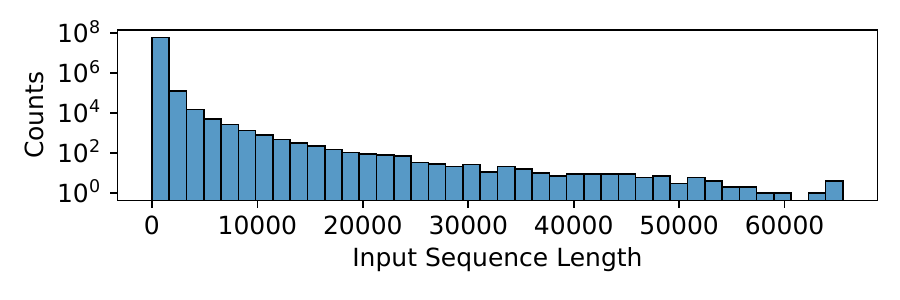}
        \caption{
        The sequence length distribution in FLANv2~\cite{Longpre2022flanv2} zero shot dataset (truncated at 65536). Y-axis is in log scale.
        }
        \label{fig:flan_seq_hist}
     \end{subfigure}
     \caption{Model inputs exhibit high sequence length variance in multi-task training.}
     \label{fig:seqlen_variance}
\end{figure}

A crucial aspect of multi-task training is the accommodation of diverse text sequence lengths across various tasks or datasets.
Tasks like summarization or information extraction usually involve a long context text as input, while only one sentence is usually used as input to simple question answering (e.g., checking the grammatical acceptability of the sentence) (Fig.~\ref{fig:example_variation}).
The average input sequence length is 977.73 tokens in the CNN/Daily Mail~\cite{hermann2015cnndm} dataset for the text summarization task, while the MNLI dataset~\cite{williams2018mnli} for textual entailment only has an average input sequence length of 51.59. 
This results in high sequence length variations in multi-task dataset mixtures (e.g., FLANv2~\cite{Longpre2022flanv2}), as shown in Fig.~\ref{fig:flan_seq_hist}.
During training, batches of uniform-length samples are usually needed to be fed into the accelerator devices (e.g., GPUs) for efficient processing.
The input sequences need to be uniformly padded to at least the largest sequence length in each mini-batch (which can be very long due to the presence of long-context tasks), resulting in excessive padding, increased memory consumption and wasted computation.

The \textit{packing}~\cite{raffel2020t5} approach has been advocated to alleviate this issue, which concatenates multiple short samples to form a long sample whose length matches the largest input sequence length.
Packing can be effective in reducing padding.
Since almost all current language models use the Transformer~\cite{vaswani2017attention} architecture, attention is computed among tokens in each long sequence during training and such attention computation is wasted among unrelated samples packed into the same sequence.
Such computation waste grows quadratically with sequence length, leading to extensive overhead in case of large sequence lengths. 
Attention computation among unrelated samples can also have negative impact on model performance~\cite{krell2021packwmask}.
Additional attention masks~\cite{wei2022flan} and adjustments of the positional embeddings~\cite{krell2021packwmask} are needed to exclude this cross-contamination effect, which complicates model implementation.

For better multi-task training efficiency, we propose a \textbf{dynamic micro-batching} approach to address the variable-length input challenges.
Pipeline parallelism is commonly adopted in LLM training~\cite{narayanan2019pipedream}: the large model is partitioned into stages deployed over multiple devices; the input mini-batch of training samples in each training iteration is partitioned into micro-batches, and the micro-batches are processed across the devices in a pipelining manner.
Our key idea is that we only need to ensure similar sequence lengths among samples within each micro-batch, but not across micro-batches, such that we can group samples accordingly, minimize padding and eliminate any unnecessary attention computation or masking among unrelated samples.

Current pipeline training systems adopt uniform micro-batch sizes, and do not efficiently support processing of micro-batches with different sequence lengths, memory consumption and execution time.
We design {\SystemName{}}, a dynamic micro-batching pipeline training framework that enables efficient multi-task model training with different input sequence lengths.
{\SystemName{}} automatically optimizes micro-batching, pipeline and communication scheduling in each training iteration. 
We make the following contributions in designing {\SystemName{}}.

\vspace{1mm}
\noindent$\triangleright$ \textbf{We devise an efficient dynamic programming-based method to optimize micro-batch construction}, that balances the trade-off between padding reduction, computation efficiency and memory consumption of the micro-batches.
Upon input of each mini-batch, we first sort the samples to minimize the sequence length difference between adjacent samples.
We then use dynamic programming to decide optimal splits of the sorted sample list into micro-batches, exploiting our cost model on per-iteration LLM training time under pipeline parallelism.

\vspace{1mm}
\noindent$\triangleright$ \textbf{We propose pipeline schedules that are robust to execution time variations of the micro-batches.}
We identify that the commonly adopted 1F1B pipeline schedule~\cite{narayanan2019pipedream} is prone to blocking (device idling during pipeline execution) under vacillating micro-batch execution time, based on the concept of safety stocks~\cite{boudoukh2001cyclicschedule}.
To mitigate the issue, we advocate adaptive scheduling which controls the injection time of micro-batches into the pipeline. We also make it memory-aware, maximizing training throughput while observing device memory limits.

\vspace{1mm}
\noindent$\triangleright$ \textbf{We design effective communication schedule to allow irregular communication patterns in our dynamic pipeline.}
Na\"ive communication schedule (i.e., sending tensors to the next pipeline stage immediately after production, and receiving them just before use) causes deadlocks in our dynamic pipelines since different processing stages of a micro-batch are no longer scheduled tightly one after another.
We perform ahead-of-time planning, scheduling both send and receive operations at the production time of each tensor, which is guaranteed to be deadlock-free.

\vspace{1mm}
We implement \SystemName{} on PyTorch~\cite{paszke2019pytorch}.
Extensive evaluation on the FLANv2 dataset~\cite{Longpre2022flanv2} reveals up to 4.39x throughput improvement when training T5 \cite{raffel2020t5}, and 3.25x when training GPT \cite{brown2020gpt3}, compared with packing-based baselines.

%% file: sections/background_motivation.tex
\section{Background and Motivation}
\subsection{Multi-task model training}

\begin{figure}[t]
    \centering
    \includegraphics[width=\linewidth]{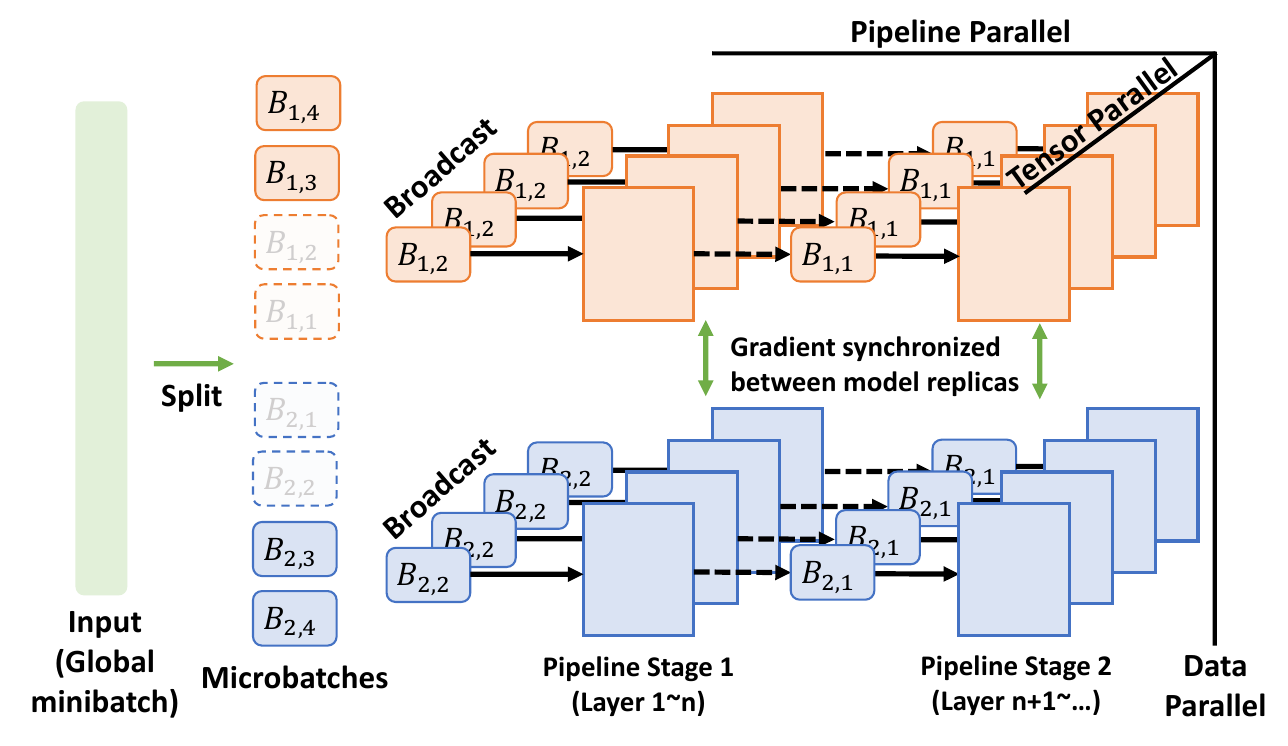}
    \caption{
    Illustration of 3D parallelism and input data partitioning. Each square block indicates (partitioned) model layers (operators) on a GPU. Different data parallel model replicas are denoted with different color (within each replica, model is partitioned through tensor and pipeline parallelism). The input global mini-batch is split into micro-batches (each containing different data). $B_{i,j}$: the $j$th micro-batch for model replica $i$.
    }
    \label{fig:3d_parallel}
\end{figure}

Modern LLMs (e.g., T5~\cite{raffel2020t5}) can perform different tasks without the need for task-specific model structures.
To perform multi-task training, we only need to produce a mixture of data from different datasets.
The mini-batches are then randomly sampled from the dataset.
The exact way to mix data (i.e., proportion of each dataset in the final mixture) can be method-dependent~\cite{raffel2020t5}.

Multi-task LLMs are commonly trained with a combination of data, tensor and pipeline parallelism (i.e., 3D parallelism) to address memory pressure induced by their large model size.
With pipeline parallelism, the model layers are partitioned among devices (stages), and a mini-batch is split into smaller micro-batches in the batch dimension.
In each training iteration, the micro-batches are sequentially executed with gradient accumulated across the micro-batches.
The forward and backward order for each device is determined by the pipeline schedule (e.g., 1F1B~\cite{narayanan2019pipedream}, where each stage executes one forward pass and one backward pass, alternatively).
Tensor parallelism shards computation within individual operators (e.g., matrix multiplication) to different devices, and is agnostic to micro-batching.
In data-parallel training, the model is replicated on each device and a different portion of the mini-batch is processed on each replica.
Fig.~\ref{fig:3d_parallel} illustrates the input partitioning with 3D parallelism.

The input samples to a multi-task model often have vastly different sequence lengths.
For efficient processing of hardware accelerator such as GPUs, samples are usually batched, forming a single input tensor.
To accommodate longer sequences, shorter samples in a input batch need to be padded to (at least) the length of the longest sequence in the batch.
Under extreme sequence length variations, the amount of padding can be substantial.
For example, na\"ive padding (padding every sample to the length of the longest sequence in a mini-batch) of samples from the FLANv2~\cite{Longpre2022flanv2} dataset leads to more than 80\% of padding tokens in a mini-batch.
Memory and computation resources are wasted processing the unused padding tokens.
While sorting (bucketing) samples by sequence length before batching can alleviate padding~\cite{ott2019fairseq}, it destroys randomness of batch sampling and may degrade model performance:
for example, after sorting, the mini-batches with long sequence lengths may only consist of samples from a small number of tasks (such as summarization);
such homogeneous batches may harm the performance of the multi-task model since it harms the model's generalizability~\cite{aghajanyan2021muppet, gottumukkala2020dynamic}.

\subsection{The current packing solution}
To alleviate the padding problem, packing is a common solution, that concatenates multiple short sequences at the sequence dimension to form a single sequence that matches a (predefined) maximum sequence length~\cite{raffel2020t5}.
Individual sequences that are longer are usually truncated.
Cross contamination between samples that are packed into the same sequence may happen, since attention is calculated across unrelated samples, affecting model prediction.
Special masks are used during self-attention~\cite{krell2021packwmask} to prevent it, which zero out the attention scores between unrelated samples.
Since the packed sequences have similar lengths, padding is greatly reduced.

\begin{figure}[t]
     \centering
     \includegraphics[width=\linewidth]{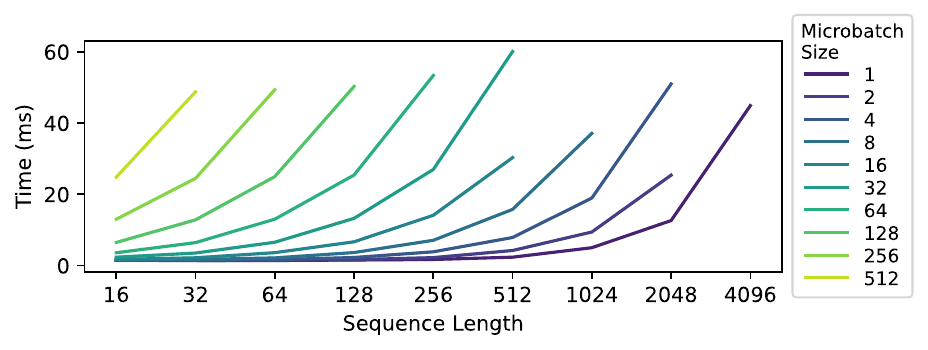}
     \caption{Computation time of a single Transformer encoder layer in T5-11B on an A100 GPU. The execution time exhibits super-linear growth with sequence length.}
     \label{fig:time_vs_seqlen}
\end{figure}

\begin{figure}[t]
     \centering
     \begin{subfigure}[b]{\linewidth}
         \centering
         \includegraphics[width=\textwidth]{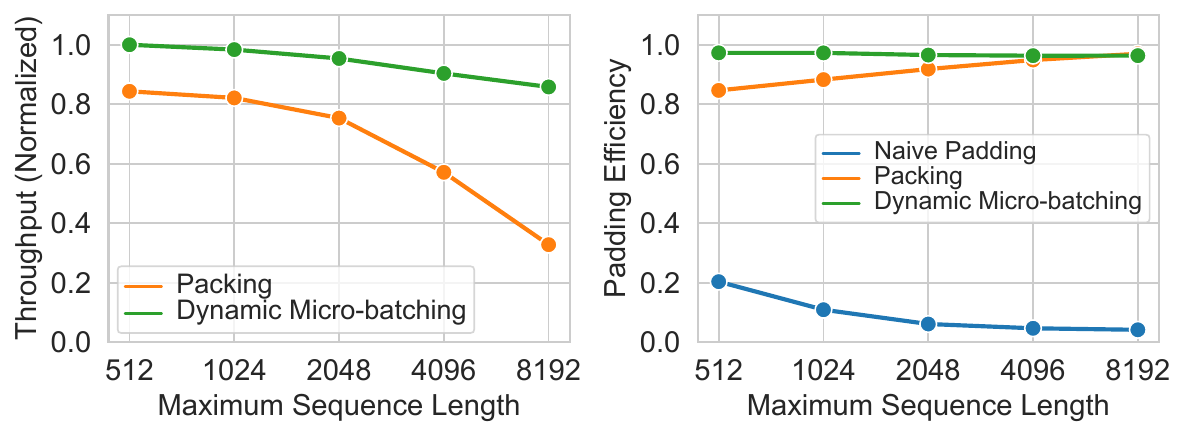}
         \caption{GPT}
         \label{fig:packing_vs_dyn_bat_gpt}
     \end{subfigure}
     \hfill
     \begin{subfigure}[b]{\linewidth}
         \centering
         \includegraphics[width=\textwidth]{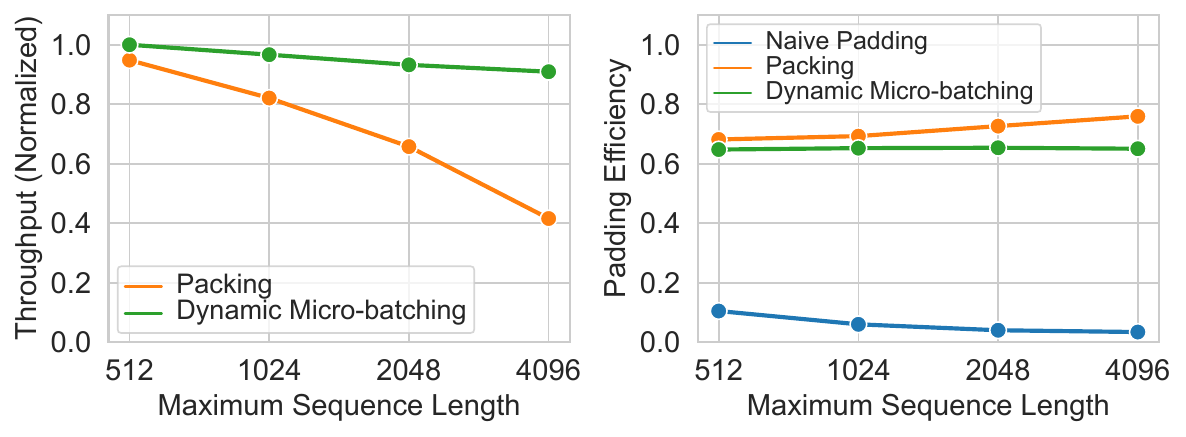}
         \caption{T5}
         \label{fig:packing_vs_dyn_bat_t5}
     \end{subfigure}
     \caption{
     GPT and T5's training performance under packing and dynamic micro-batching.
     }
     \label{fig:dyn_vs_pack_simulation}
\end{figure}

However, computation overhead of packed sequences may increase substantially with the maximum sequence length.
The computation complexity of self-attention increases quadratically with sequence length~\cite{vaswani2017attention}.
Fig.~\ref{fig:time_vs_seqlen} shows super-linear increase in computation time of a Transformer layer (an LLM is often a stack of Transformer layers) with the sequence length.
Training throughput of the whole model is given in Fig.~\ref{fig:dyn_vs_pack_simulation}, when training GPT~\cite{brown2020gpt3} and T5~\cite{raffel2020t5} on the FLANv2 dataset.
We observe more than 50\% throughput decrease when the maximum sequence length increases from 512 to 8192, while the total number of non-padding tokens only increases by 13.2\% (due to less truncation).

We advocate dynamic micro-batching that adapts the number of micro-batches and micro-batch sizes in each training iteration according to the input data, to efficiently tackle the variable sequence length problem.
By grouping samples with similar sequence lengths into the same micro-batch, we reduce the amount of padding needed without introducing unnecessary attention computation as in packing.
Fig.~\ref{fig:dyn_vs_pack_simulation} gives preliminary comparison of the dynamic batching approach (with micro-batches split using our dynamic programming method) with packing under the same settings.
The padding efficiency is computed by dividing the non-padding tokens by the total number of tokens processed (padding and non-padding).
Dynamic micro-batching achieves comparable padding efficiency as packing and better training throughput, which only slightly drops when the maximum sequence length increases.

\subsection{Challenges of dynamic micro-batching}

\begin{figure}[t]
     \centering
     \begin{subfigure}[b]{\linewidth}
         \centering
         \includegraphics[width=\textwidth]{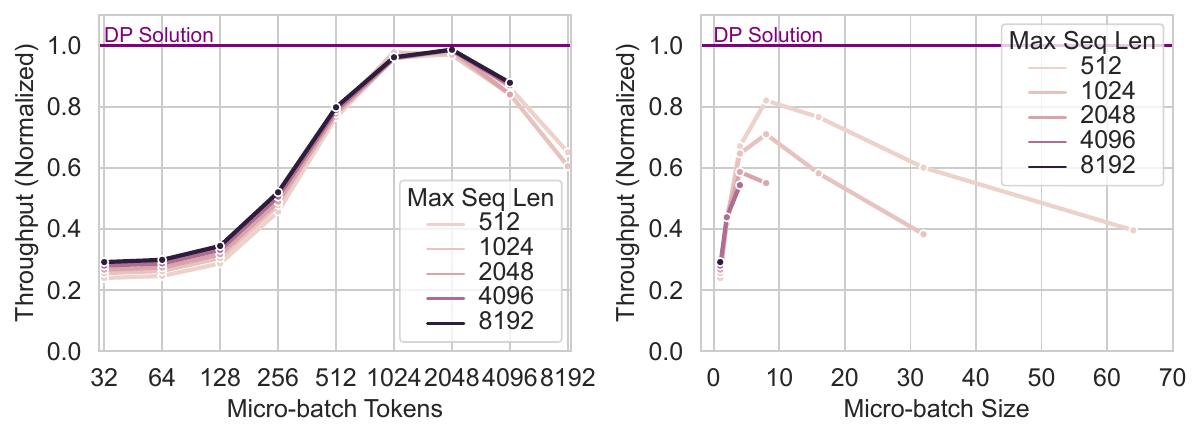}
         \caption{GPT}
         \label{fig:batching_methods_gpt}
     \end{subfigure}
     \hfill
     \begin{subfigure}[b]{\linewidth}
         \centering
         \includegraphics[width=\textwidth]{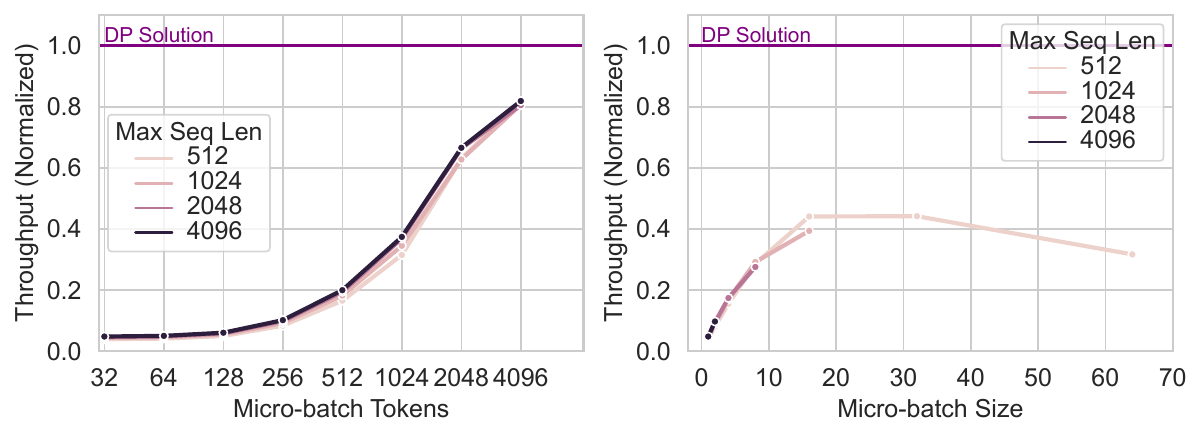}
         \caption{T5}
         \label{fig:batching_methods_t5}
     \end{subfigure}
     \caption{Training performance of GPT and T5 under different micro-batching methods. Left figures show results for token-based micro-batching, and right figures are for micro-batching using 
     fixed micro-batch size. All throughput values are normalized over that achieved by our dynamic programming (DP)-based micro-batching method.}
     \label{fig:micro_batching_methods}
\end{figure}

$\triangleright$ \textbf{No principled way to split training mini-batches into micro-batches of different sequence lengths.}
Most current pipeline training systems use micro-batches of exactly the same shape: the same number of samples per micro-batch (i.e., the same micro-batch size) and the same sequence length among samples in the micro-batches (padded or packed sequences in case of different sequence lengths).
Other possible methods include generating micro-batches of the same token count, so that there are fewer samples in micro-batches of larger sequence lengths.
Fig.~\ref{fig:micro_batching_methods} shows training throughput under the two micro-batching methods.
Using a uniform micro-batch size leads to out-of-memory errors (OOMs) when the micro-batch size increases and the maximum sequence length is large;
when the maximum sequence length is small, the performance first improves due to the increase in computation efficiency, and then drops because of more padding at larger micro-batch sizes.
Equal token count-based micro-batching achieves much better training throughput, while still experiencing OOM before reaching the highest throughput during T5 training.

We observe that the choice of micro-batch size or the token number in the two methods greatly affects the training throughput.
We design an efficient dynamic programming-based algorithm to decide optimal micro-batching in each training iteration, that strikes a good trade-off between padding efficiency, computation efficiency and memory consumption.

\begin{figure}[t]
     \centering
      \begin{subfigure}[b]{\linewidth}
         \centering
         \includegraphics[width=\textwidth]{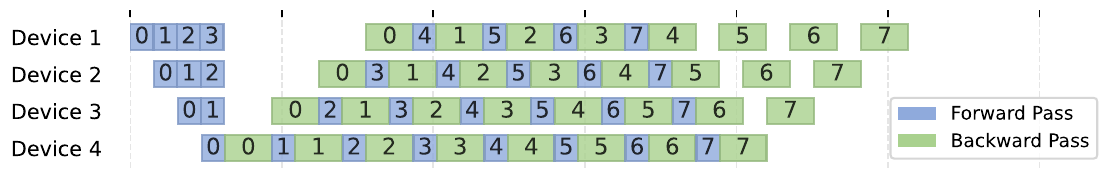}
         \caption{1F1B (uniform micro-batches)}
         \label{fig:sched_compare_1f1b_uniform}
     \end{subfigure}
     \hfill
      \begin{subfigure}[b]{\linewidth}
         \centering
         \includegraphics[width=\textwidth]{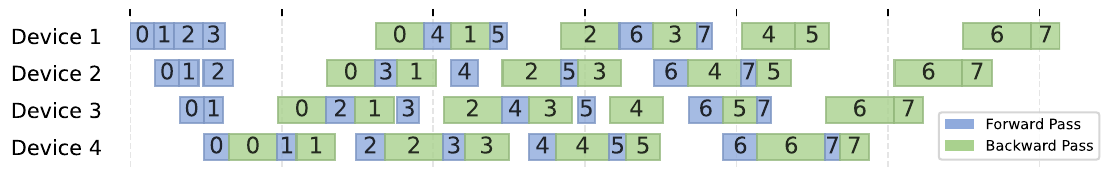}
         \caption{1F1B (dynamic micro-batches)}
         \label{fig:sched_compare_1f1b_heter}
     \end{subfigure}
     \hfill
      \begin{subfigure}[b]{\linewidth}
         \centering
         \includegraphics[width=\textwidth]{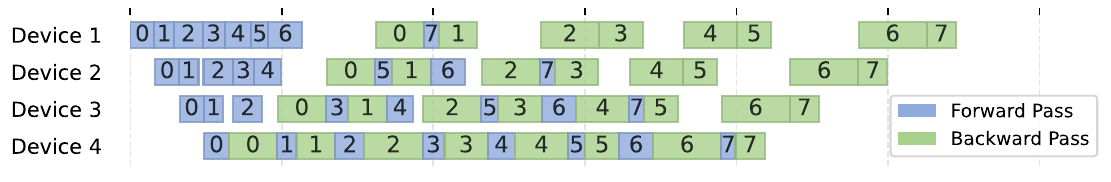}
         \caption{Adaptive Schedule}
         \label{fig:sched_compare_cyclic}
     \end{subfigure}
     \hfill
      \begin{subfigure}[b]{\linewidth}
         \centering
         \includegraphics[width=\textwidth]{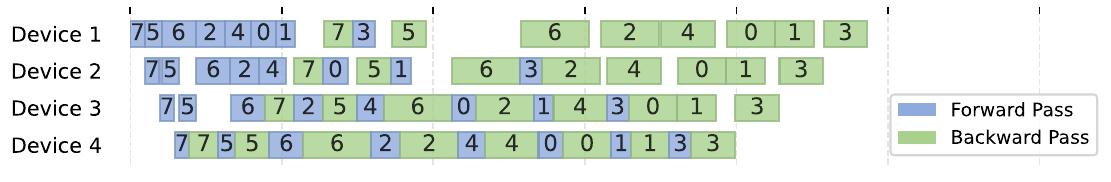}
         \caption{Adaptive Schedule \& Micro-batch Reordering}
         \label{fig:sched_compare_cyclic_reorder}
     \end{subfigure}
     \caption{Pipeline training under dynamic micro-batching (except \ref{fig:sched_compare_1f1b_uniform}) with different pipeline schedules. 
     }
     \label{fig:schedule_comparison}
\end{figure}

\vspace{1mm}
\noindent$\triangleright$ \textbf{No efficient pipeline schedules for micro-batches of diverse execution times.}
Most existing pipeline schedules (e.g., 1F1B~\cite{narayanan2019pipedream}) assume identical execution time of micro-batches, and schedule micro-batch processing over consecutive stages tightly one after another (Fig.~\ref{fig:sched_compare_1f1b_uniform}).
In this way, any variation in micro-batches' execution time may cause blocking (computation/communication waiting for communication/computation), creating more ``bubbles'' in the training pipeline (Fig.~\ref{fig:sched_compare_1f1b_heter}).

\begin{figure}[t]
    \centering
    \includegraphics[width=0.8\linewidth]{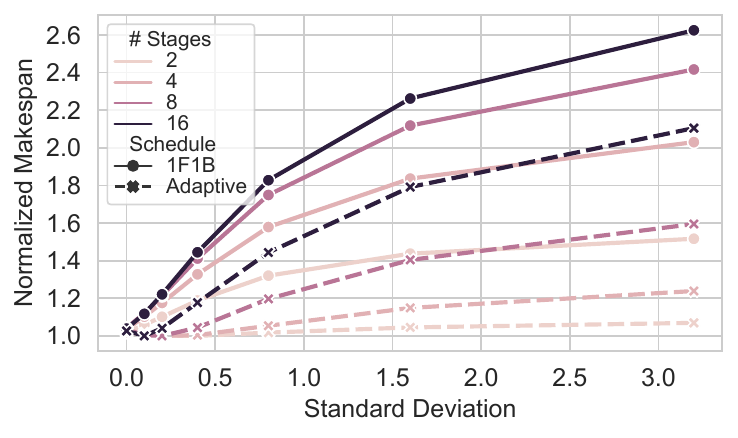}
    \caption{
    Per-iteration training time (makespan) of different pipeline schedules under different variation levels of micro-batch execution time. X axis is the standard deviation of the introduced variation (zero-mean Gaussian). The makespan is normalized over the no variation case.
    }
    \label{fig:schedule_under_variation}
\end{figure}

To further quantify the effect of variable micro-batch execution time on 1F1B schedule, we randomly disturb the execution time of the micro-batches (assumed to be uniform originally) by noises from a zero-mean Gaussian distribution.
Fig.~\ref{fig:schedule_under_variation} shows that the per-iteration training time with 1F1B schedule grows rapidly as the variation level increases, especially when there are more pipeline stages.

We present a scheduling algorithm that is more robust to dynamic micro-batches, as shown in Fig.~\ref{fig:schedule_under_variation}. 
However, to achieve a higher throughput, it consumes more memory than 1F1B.
We make the algorithm memory-aware, so it achieves higher throughput than 1F1B when memory is abundant, while automatically limiting memory consumption under high memory pressure.

\begin{figure}[t]
     \centering
     \begin{subfigure}[b]{\linewidth}
        \centering
        \includegraphics[width=\textwidth]{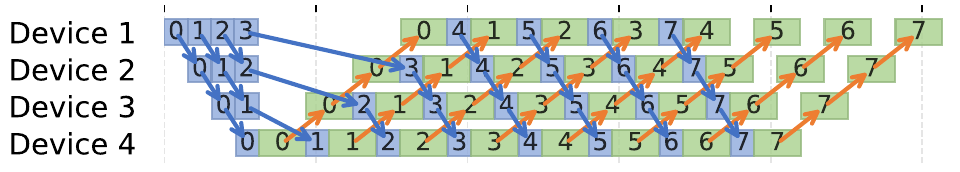}
        \caption{Regular communication pattern of 1F1B schedule.}
        \label{fig:comm_pattern_1f1b}
     \end{subfigure}
     \hfill
     \begin{subfigure}[b]{\linewidth}
         \centering
         \includegraphics[width=\textwidth]{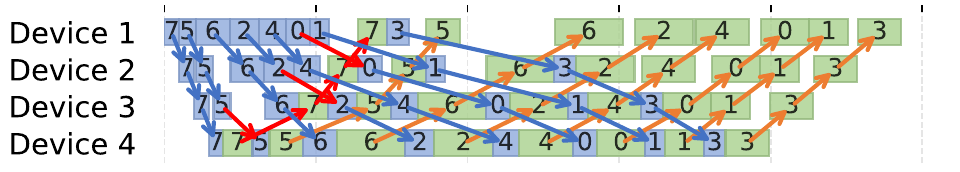}
         \caption{Irregular and dynamic communication pattern of dynamic pipelines. Red arrows indicate communications which can trigger deadlock if not scheduled correctly (only one shown for each pair of devices).}
         \label{fig:comm_pattern_dynamic}
     \end{subfigure}
     \caption{Difference in communication pattern between 1F1B (with uniform micro-batches) and \SystemName{} schedules. Green (blue) blocks denote backward (forward) computation. Blue and orange arrows indicate the communication of activations during forward pass and gradients during the backward pass, respectively.
     }
     \label{fig:communication_pattern}
\end{figure}

\vspace{1mm}
\noindent$\triangleright$ \textbf{Improper communication order between pipeline stages may lead to deadlocks in dynamic pipelines.}
Current pipeline systems send an intermediate tensor to the next stage right after its production, and launch the receive operation of the tensor just before it is used, when applying 1F1B schedule.
Since there is no gap between consecutive execution stages of a micro-batch in 1F1B schedule, the send and receive ops naturally align in time.
As shown in Fig.~\ref{fig:comm_pattern_1f1b}, each crossing of arrows (a pair of sends in reverse direction) can be implemented by a fused communication operator.
However, our pipeline schedule is different for each iteration and can produce irregular communication patterns where execution of consecutive stages of the same micro-batch are scheduled far apart (Fig.~\ref{fig:comm_pattern_dynamic}), causing deadlocks for na\"ive communication schedule (i.e., start sending whenever the result of a stage is ready and start receiving whenever previous stage's result is needed).
For example, the uppermost red arrows in Fig.~\ref{fig:comm_pattern_dynamic} shows device 1 sending the activation of micro-batch 0 to device 2, while at the same time, device 2 is trying to send the gradient of micro-batch 7 to device 1.
However, under na\"ive schedule, device 1 will continue to send the activation of micro-batch 1 to device 2 before launching a corresponding receive for micro-batch 7.
Since only one communication operation can happen between each pairs of devices (required by libraries like NCCL~\cite{nvidia2023nccl}), this creates a communication order mismatch thus can result in deadlocks (fusing of communication ops like in 1F1B scheduling is also infeasible due to the extra sending of the activation of micro-batch 1).

We reorder the send and receive operations to resolve deadlocking.
For example, we can make Device 1 receive micro-batch 0's activation before it sends micro-batch 7's gradient, and make Device 1 receive micro-batch 7's gradient before it sends micro-batch 1's activation. 
Given a pipeline schedule, we schedule both send and receive operators together when an intermediate tensor is produced, ensuring the communication order is consistent across different stages.

%% file: sections/overview.tex
\section{\SystemName{} Overview}
\label{sec:overview}

\begin{figure}[t]
    \centering
    \includegraphics[width=0.9\linewidth]{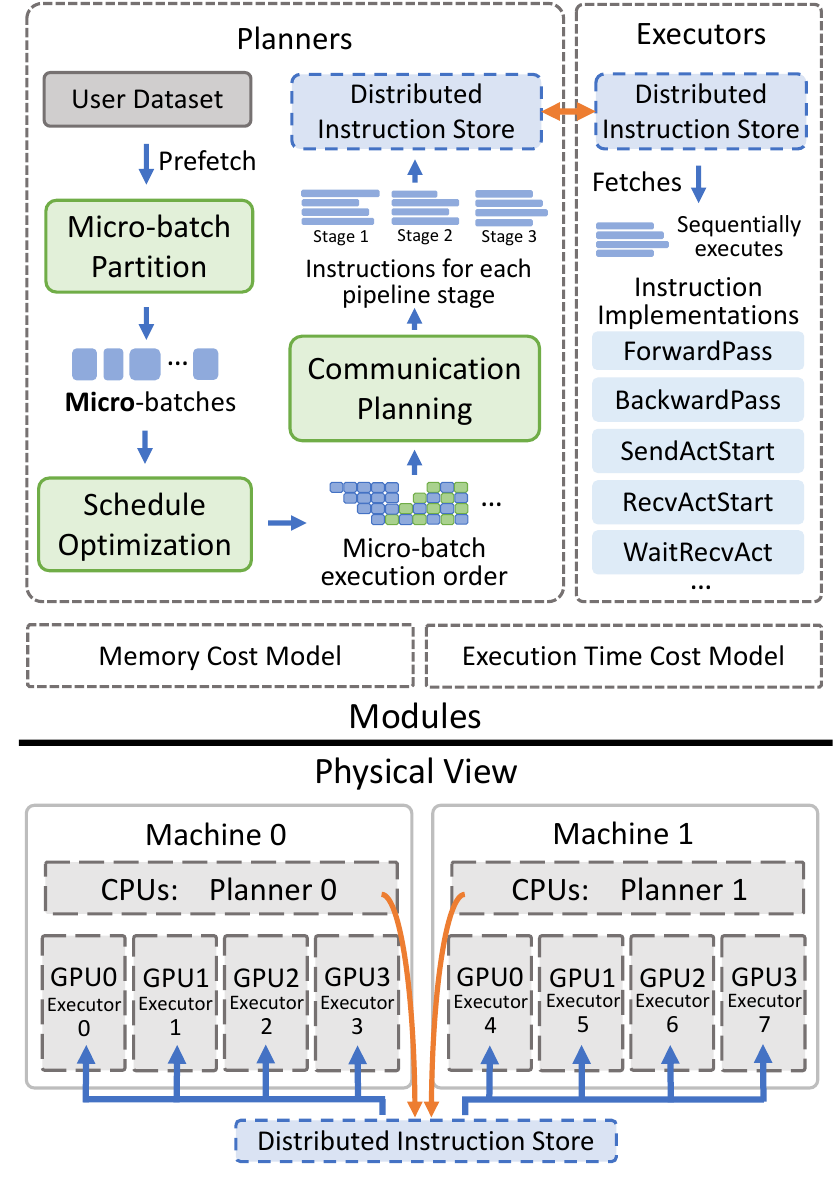}
    \caption{
    System architecture of \SystemName{}. 
    }
    \label{fig:overview}
\end{figure}

We propose {\SystemName{}} to enable efficient pipeline training of multi-task models with dynamic micro-batching.
{\SystemName{}} comprises of two main modules:
(1) \textit{Planners} that run on CPUs, perform optimization and generate \textit{execution plans} for each training iteration;
(2) \textit{Executors} that retrieve and execute the assigned execution plans on the GPUs.
A system overview of {\SystemName{}} is given in Fig.~\ref{fig:overview}.
We detail the components and terminologies used in our design as follows.

\vspace{1mm}
\noindent\textbf{Execution plans} specify micro-batch splitting, pipeline execution schedule, the communication order and the shape of all communicated tensors on each executor (GPU).
They are represented as sequences of \textit{pipeline instructions}, following the design principle of DeepSpeed~\cite{deepspeed}.
The \textit{pipeline instructions} include \textit{ForwardPass, BackwardPass}, which executes the forward/backward computation for a micro-batch, \textit{SendAct, RecvAct, SendGrad, RecvGrad}, which sends/receives model activation/gradients.
These \textit{pipeline instructions} abstract key operations (ops) in pipeline-parallel training.
We further split every type of communication into two conjugate pipeline instructions:
a \textit{Start} op (e.g., \textit{RecvActStart}), which launches the communication into an asynchronous GPU communication stream,  and a \textit{Wait} op (e.g., \textit{WaitRecvAct}), which adds a dependency between the communication stream and the computation stream, allowing computation to wait for the result of communication.
The instruction abstraction allows flexible pipeline scheduling and communication planning.

\vspace{1mm}
\noindent\textbf{Planners} pre-fetch training data from user-provided dataset.
For each mini-batch of training samples, a planner splits the mini-batch into micro-batches using our dynamic programming algorithm (\S\ref{sec:micro-batching}).
It then generates an optimized pipeline execution schedule (\S\ref{sec:pipeline_schedule}) and decides the appropriate execution order of communication between pipeline stages (\S\ref{sec:comm_planning}), along with the shapes of the communicated tensors.
All the above decisions are compiled into an \textit{execution plan} and pushed to a distributed instruction store in the host memory of one of the machines (e.g., machine 0), ready to be fetched by corresponding executors.

\vspace{1mm}
\noindent\textbf{Executors} retrieve execution plans from the instruction store and executes the pipeline instructions in the order specified in the execution plans using the underlying deep learning framework (e.g., Megatron-LM~\cite{narayanan2021efficientmlm} and PyTorch~\cite{paszke2019pytorch}).

\vspace{1mm}
\noindent\textbf{Cost models} estimate the execution time and memory consumption of a single layer of the model executing a micro-batch under different micro-batch sizes and sequence lengths on a single GPU.
They are used to guide all decisions in the planners.
To construct these cost models, we run memory consumption and execution time profiling for both forward and backward passes under different combinations of micro-batch size and sequence length at power-of-two intervals (e.g., micro-batch size of $1,2,4$, etc, and sequence length of $32, 64, 128$, etc).
For training with only data and pipeline parallelism, single-GPU profiling is sufficient since the communication cost is constant (under different micro-batch size and sequence lengths) for data parallelism and very small for pipeline parallelism.
Tensor parallelism profiling runs multi-GPUs to capture the significant communication cost.
Linear interpolation is used to bridge the gaps between sampled data points.
We show that this simple cost modeling suffices to provide good estimation of execution time and peak memory consumption for training iterations in Sec.~\ref{sec:cost_model_accuracy}.

To hide the plan generation overhead, we overlap model execution and the execution plan generation of future iterations in {\SystemName{}}, as planners and executors run on different hardware resources.
We exploit the abundance of CPU cores to parallelize plan generation on a machine. When the training is run on multiple machines, \SystemName{} distributes execution plan generation of distinct training iterations to different machines.

%% file: sections/microbatch_partition.tex
\section{Micro-batch Construction}
\label{sec:micro-batching}
 
In each training iteration, micro-batching samples of different sequence lengths should attend to the trade-offs among padding, computation time, memory consumption and pipeline bubble size, for throughput maximization without OOM.
We present an algorithm to optimize micro-batch partitioning.

\begin{figure}[t]
    \centering
    \includegraphics[width=\linewidth]{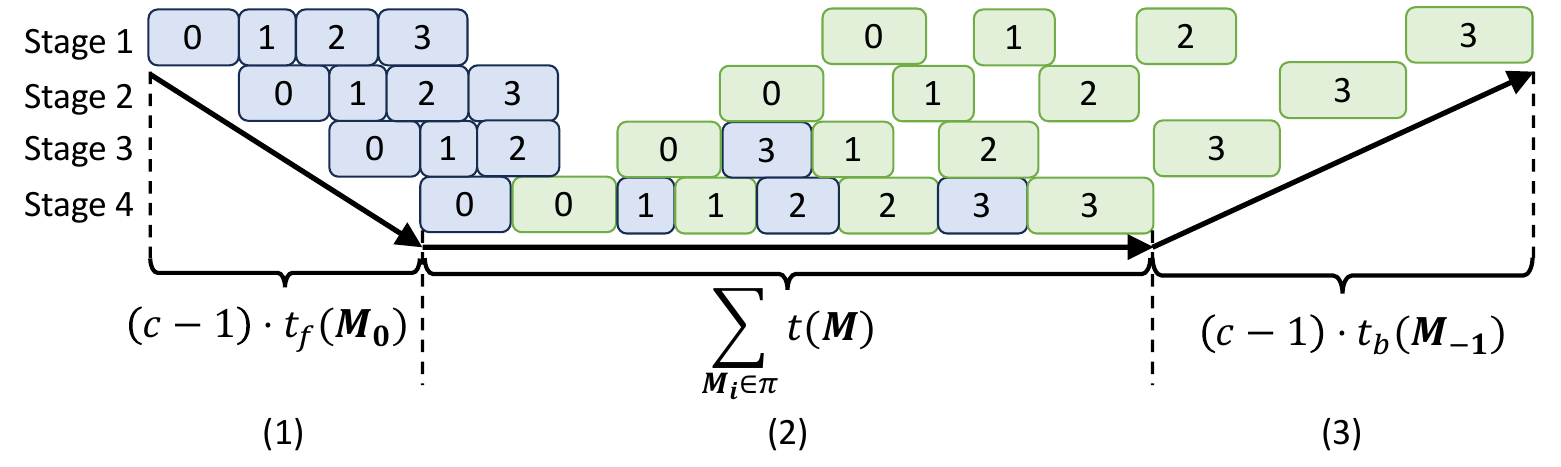}
    \caption{
    Approximation of pipeline execution time. $c$: the number of pipeline stages, $c=4$ in this example. $\mathbf{M}_0, \mathbf{M}_{-1}$: the first and the last micro-batch. $t_f(\mathbf{M}), t_b(\mathbf{M})$: the forward and backward execution time of micro-batch $\mathbf{M}$. $t_f(\mathbf{M}) + t_b(\mathbf{M}) = t(\mathbf{M})$.
    }
    \label{fig:appox_pipeline_cost}
\end{figure}

\paragraph{Model the iteration time under pipeline parallelism.}
We group a set of $N$ input sequences (samples), $\mathbf{S}$, in the current training iteration into a set of micro-batches, $\bm{\pi}=\{\mathbf{M_1, M_2,\cdots,M_m}\}$, where $\mathbf{M_i} \subseteq \mathbf{S}$ represents a micro-batch and $\mathbf{M_i}$'s are disjoint.
Let $t_f(\mathbf{M_i})$ ($t_b(\mathbf{M_i})$) be the forward (backward) pass execution time of micro-batch $i$ obtained from the cost model, and let $t(\mathbf{M_i}) = t_f(\mathbf{M_i}) + t_b(\mathbf{M_i})$.
Let $c$ be the number of pipeline stages.
The execution time of the entire pipeline can be modelled in three parts (Fig.~\ref{fig:appox_pipeline_cost}):
(1) the time for the forward pass of the first micro-batch ($\mathbf{M}_0$) to reach the last stage, $(c-1) \cdot t_f(\mathbf{M}_0)$;
(2) the execution time of all micro-batches on the last stage, $\sum_{\mathbf{M}_i \in \mathbf{\pi}} t(\mathbf{M})$;
(3) the time for the backward pass of the last micro-batch ($\mathbf{M}_{-1}$) to reach the first stage, $(c-1) \cdot t_f(\mathbf{M}_{-1})$. 
However, under dynamic micro-batching where the execution time of micro-batches differs, $t_f(\mathbf{M}_0)$ and $t_b(\mathbf{M}_{-1})$ is determined by the exact pipeline schedule which we do not know in advance.
Therefore we model $t_f(\mathbf{M}_0)$ and $t_b(\mathbf{M}_{-1})$ using the execution time of longest micro-batch, so the total execution time of part (1) and (3) can be approximated as $(c-1)\cdot \max\{t(\mathbf{M_i}) | \mathbf{M_i} \in \bm{\pi}\}$.
The iteration time is thus $t_{iter} = (c-1)\cdot \max\{t(\mathbf{M_i}) | \mathbf{M_i} \in \bm{\pi}\} + \sum_{\mathbf{M_i}\in\bm{\pi}}{t(\mathbf{M_i})}$.
We seek to derive the best micro-batch assignment $\bm{\pi}$ that minimizes the iteration time, i.e., maximizes the training throughput: 

\begin{equation}
\min_{\bm\pi}~~ \left\{(c-1)\cdot \max\{t(\mathbf{M_i}) | \mathbf{M_i} \in \bm{\pi}\} + \sum_{\mathbf{M_i}\in\bm{\pi}}{t(\mathbf{M_i})}\right\} \label{eqn:t_iter} \tag{Eq1}
\end{equation}

\vspace{1mm}
\noindent\textbf{Determine the order of samples.}
The problem of assigning samples into disjoint sub-sets (micro-batches) while optimizing an objective (throughput) belongs to the family of set partitioning problems (SPP), which is NP-hard~\cite{karp2010reducibility}.
We simplify the problem by ordering the samples first and then grouping consecutive samples into micro-batches using a dynamic programming (DP) approach.

For sample ordering, a natural intuition is that \textit{to minimize padding, micro-batches should contain samples with similar sequence lengths}.
For decoder-only models (e.g. GPT~\cite{brown2020gpt3}), sorting the samples according to their sequence lengths suffices.
For encoder-decoder models like T5~\cite{raffel2020t5} with multiple input sequences (i.e. a input sequence processed by encoder, and a target sequence fed into the decoder), we can sort the samples first by the length of the input sequence and then by the target sequence.
Alternatively, we can take the pair of input and target sequence lengths as a 2D point, and find a visiting order that minimizes the sum (or maximum) of distances between adjacent points.
This can be solved by an off-the-shelf Travelling Salesmen Problem solver.
We compare the two methods to order samples in Sec.~\ref{sec:ablation}.

\paragraph{Partition ordered samples with dynamic programming.}
Now we have an ordered list of samples $\mathbf{S}=[\mathbf{s_1, s_2, \cdots, s_N}]$.
We construct a DP algorithm to optimally partition the list.
Let $\bm{\pi}^{*}_{\mathbf{S[:n]}}$ represent the optimal partition of the first $n$ samples in $\mathbf{S}$, which minimizes the total execution time of the resulting micro-batches. Let $f(n;t_{max}) = \sum_{\mathbf{M_i} \in \bm{\pi}^{*}_{\mathbf{S[:n]}}}t(\mathbf{M_i})$, where $t_{max}$ denotes the maximum micro-batch execution time.
We have 
\begin{multline}
f(n;t_{max}) = \min_{1\leq i \leq n-1}\{f(i; t_{max}) + \\ 
t(\mathbf{M_{S[i+1:n]}}) | t(\mathbf{M_{S[i+1:n]}}) \leq t_{max}\} \label{eq:dp_state_transition} \tag{Eq2}
\end{multline}
where $\mathbf{M_{S[i+1:n]}}$ denotes the micro-batch that is constructed from samples $\mathbf{s_{i+1}, s_{i+2},\cdots,s_{n}}$.
To find the best micro-batch partitions minimizing 
(\ref{eqn:t_iter}), we only need to find $t_{max}$ that minimizes $(c-1)\cdot t_{max} + f(N; t_{max})$.
There are $O(N^2)$ unique possible $t_{max}$ values, since there are at most $\frac{N(N+1)}{2}$ ways to construct a single micro-batch by consecutively partitioning $\mathbf{S}$.
For each $t_{max}$, finding $f(N; t_{max})$ takes $O(N^2)$ steps. 
Therefore, the computation complexity of this DP approach is $O(N^4)$. 
With input sequences in multi-task model training, many $t_{max}$ values are very similar to each other.
We may greatly speed up the DP algorithm by sampling $t_{max}$ at fixed intervals (in our evaluation, we only consider possible $t_{max}$ values $5us$ apart from each other).

\paragraph{Balance data parallel model replicas.}
The above algorithm splits an input mini-batch into micro-batches for execution in a single pipeline. 
When the pipeline training is combined with data parallelism, 
the micro-batches should also be distributed among different data-parallel model replicas, balancing the execution time between all pipelines.
We extend our micro-batching algorithm to handle hybrid data and pipeline parallel training.
Let $\bm{\pi_d} \subseteq \bm{\pi}$ be the collection of micro-batches for model replica $d$.
The iteration time under hybrid data and pipeline parallelism becomes 
$$t^{dpp}_{iter} = \max_{d}\left\{(c-1)\cdot \max\{t(\mathbf{M_i}) | \mathbf{M_i} \in \bm{\pi_d}\} + \sum_{\mathbf{M_i}\in\bm{\pi_d}}{t(\mathbf{M_i})}\right\},$$
which denotes the maximum execution time across all data parallel model replicas. 
We minimize its upper bound 
$$(c-1)\cdot \max\{t(\mathbf{M_i}) | \mathbf{M_i} \in \bm{\pi}\} + \max_d\left\{\sum_{\mathbf{M_i}\in\bm{\pi_d}}{t(\mathbf{M_i})}\right\}$$
in our micro-batching.
The first term is the same as that in (\ref{eqn:t_iter}), and the second term is the maximum total micro-batch execution time among model replicas.
Minimizing the second term is not easy, since it requires solving another subset partition problem, which is NP-hard~\cite{karp2010reducibility}.
We approximate the second term using the tight lower bound $\frac{1}{|\mathbf{D}|}\sum_{\mathbf{M_i}\in\bm{\pi}}{t(\mathbf{M_i})}$, which is achieved when the total micro-batch execution time is equal across all model replicas.
Then, the approximated objective to minimize becomes 
$$\min_{\bm\pi}~~~\left\{(c-1)\cdot \max\{t(\mathbf{M_i}) | \mathbf{M_i} \in \bm{\pi}\} + \frac{1}{|\mathbf{D}|}\sum_{\mathbf{M_i}\in\bm{\pi}}{t(\mathbf{M_i})}\right\}.$$
Comparing it with (\ref{eqn:t_iter}), the only difference lies in the constant $\frac{1}{|\mathbf{D}|}$ in the second term.
Therefore, we can optimize this objective using the same DP algorithm:
we first solve for micro-batch partitioning to minimize the above objective, and then partition resulting micro-batches among data parallel replicas to minimize $\max_d\left\{\sum_{\mathbf{M_i}\in\bm{\pi_d}}{t(\mathbf{M_i})}\right\}$, by (approximately) solving the subset partition problem using the Karmarkar–Karp algorithm~\cite{karmarkar1982differencing}.

\paragraph{Limit memory consumption.}
In pipeline parallel training, the peak memory consumption (in 1F1B scheduling) is determined by the maximum accumulated activation during a sliding window of $c$ micro-batches, since on the $n$th device, it executes $c + 1 - n$ forward passes, followed by regular backward-forward cycles.
Such a sliding window introduces dependency between micro-batching decisions, which destroys the optimal substructure property of DP.
Therefore, we resort to limiting the memory consumption of every micro-batch $\mathbf{M_{S[i+1:n]}}$ during DP, i.e., only considering $\mathbf{M_{S[i+1:n]}}$s that do not violate memory limit in (\ref{eq:dp_state_transition}).
The per-micro-batch memory limit is related to the pipeline schedule. 
In 1F1B schedule, the per-micro-batch memory limit is set to $\frac{1}{c}$ of the device memory limit. In the next section, we show different schedules with different such factors ranging from 1 to $\frac{1}{m}$ (m is the number of micro-batches).

%% file: sections/schedule_optimization.tex
\section{Pipeline Execution Schedule}
\label{sec:pipeline_schedule}

We analyze pipeline execution performance under dynamic micro-batching and propose pipeline schedules for better throughput under non-uniform micro-batch execution time.

\begin{figure}[t]
     \centering
     \begin{subfigure}[b]{\linewidth}
        \centering
        \includegraphics[width=\linewidth]{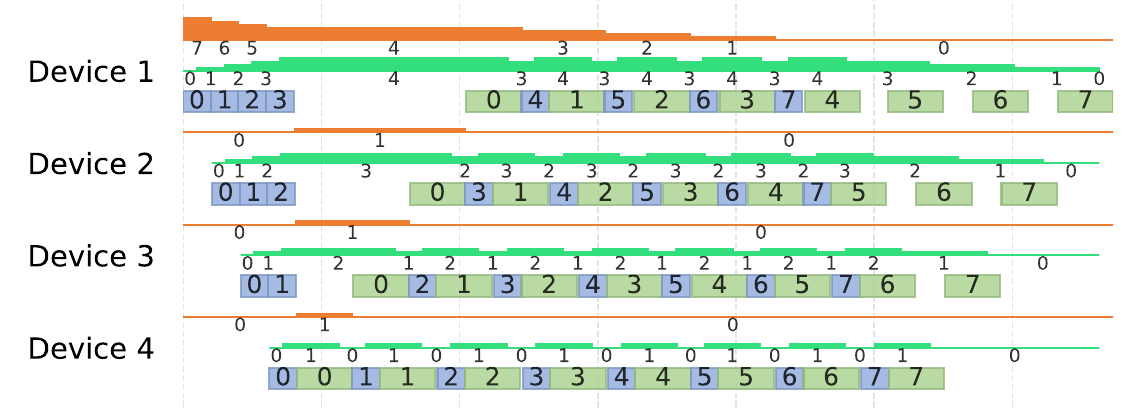}
        \caption{1F1B schedule, with zero safety stock across the steady state.}
        \label{fig:safetystocks_1f1b}
     \end{subfigure}
     \hfill
     \begin{subfigure}[b]{\linewidth}
         \centering
         \includegraphics[width=\textwidth]{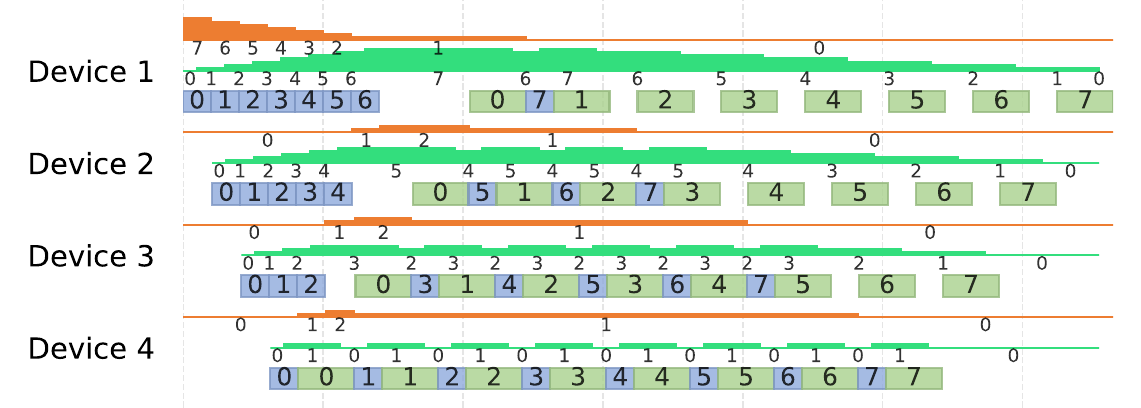}
         \caption{Adaptive schedule which injects 7 micro-batches in the beginning, with 1 safety stock across the steady state.}
         \label{fig:safetystocks_cyclic}
     \end{subfigure}
     \hfill
     \begin{subfigure}[b]{\linewidth}
         \centering
         \includegraphics[width=\textwidth]{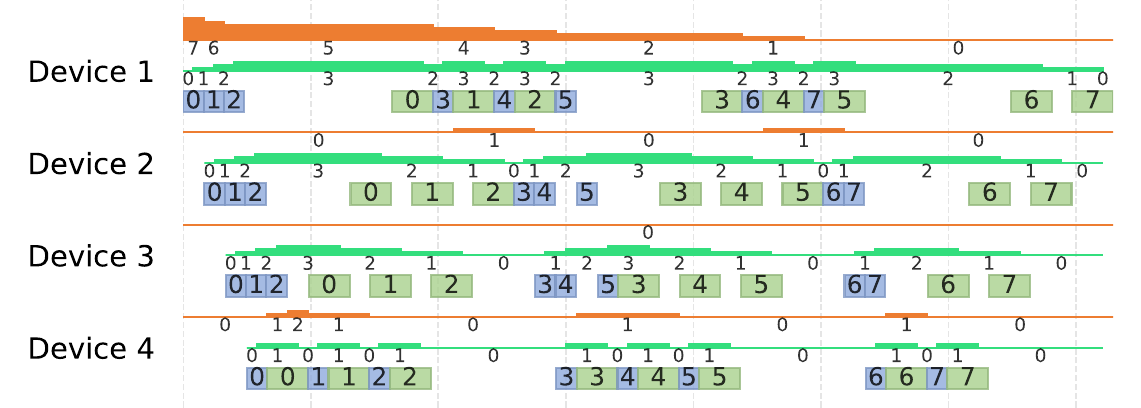}
         \caption{Adaptive schedule with peak memory limited to 3 micro-batch activations. Micro-batch 3 and 6 are delayed due to memory limit.}
         \label{fig:safetystocks_cyclic_memlimit}
     \end{subfigure}
     \caption{The trade-off between safety stock size (orange row for each device) and memory consumption (green row) under different pipeline schedules.}
     \label{fig:safetystocks}
\end{figure}

\paragraph{Analysis of pipeline execution performance.}
Let op $f_{i,j}$ ($b_{i,j}$) denote the forward (backward) computation of micro-batch $i$ on the $j$th device.
We associate a virtual buffer with each device, containing the ops ready for execution on it (i.e., whose proceeding forward/backward stages have been completed).
Ops are removed from this buffer when its execution on the device starts.
Whenever device $j$ has executed an op $f_{i, j}$ or $b_{i,j}$, the op of the next stage, $f_{i, j+1}$ ($b_{i, j}$ for the final forward stage) or $b_{i, j-1}$ is added to the buffer of the corresponding device ($j+1$ or $j-1$).
Ops in this buffer are called safety stocks of that device, in the scheduling literature~\cite{boudoukh2001cyclicschedule}.
To prevent device idling, it is essential to maintain non-empty safety stocks when the device has executed an op and is ready for the next.
In 1F1B scheduling shown in Fig.~\ref{fig:safetystocks_1f1b}, only the first stage has 7 safety stocks initially since all micro-batches are ready for execution (no dependencies on prior stages); 
its safety stock gradually depletes as more micro-batches are executed.
For all other stages, the number of safety stocks is zero throughout the steady states (the period where forwards and backwards are closely packed and follows strict one-forward-one-backward order).
This is because 1F1B schedules consecutive stages closely together without any gap time in between.
Whenever the previous stage finishes executing an micro-batch (and thus the added to the safety stock of the current stage), the current stage will immediately start the computation of this micro-batch, resulting in a net zero change in safety stock level.
With zero safety stocks, any deviation in micro-batch execution time will result in device idling.

\paragraph{Schedules robust for dynamic micro-batching.}
We seek a pipeline execution schedule that maintains more safety stocks at each device. 
The problem of micro-batch execution scheduling can be viewed as a special type of the re-entrant flow shop problem~\cite{graves1983reentrant} (i.e., scheduling jobs onto machines where each job follow the same process route through the machines; a machine can be used more than once by a job), since all our micro-batches passes through the devices following the same routine (perform forward computation once on each stage, then follow the reverse route backward).

Cyclic scheduling is an algorithm that has demonstrated commendable performance in solving re-entrant flow shop problems~\cite{boudoukh2001cyclicschedule}.
Under cyclic scheduling, execution on each device is divided into cycles; in each cycle, each device tries to execute exactly one forward pass and one backward pass of any micro-batch.
Each device maintains two buffers of ready ops (forward and backward), and fetches an op to execute from each buffer in each cycle.
If no ops are available, the corresponding forward or backward pass is skipped.
Like 1F1B schedule, cyclic schedule will interleave forward and backward pass of micro-batches during the course of scheduling.
However, unlike 1F1B schedule which fixes the execution order of all micro-batches regardless of their execution time and memory consumption, cyclic schedule provides us with a systematic way of controlling when micro-batches should be injected into the pipeline during the scheduling process (We can mark all micro-batches as not ``ready'' on the first stage during initialization. To inject a micro-batch, we insert it into the buffer of ready forward ops on the first device).
We refer to such micro-batch-injection-regulated cyclic schedules as adaptive scheduling, since the injection time can be adjusted adaptively for different input micro-batches.

Micro-batch injection time in turn affects the level of safety stocks.
For example in Fig.~\ref{fig:safetystocks_cyclic}, if we inject more micro-batches into the pipeline at the beginning, we raise the number of safety stock to one at each device during the steady state.
This means at least one micro-batch is ready for execution for each device, therefore, a device will not idle even if the previous stage does not produce the activations in time (e.g., when executing a large micro-batch).
The increase in safety stock level leaves room for variations in execution time of the micro-batches.

\paragraph{Optimizing trade-off between time and memory.}
Injecting more micro-batches also increases the memory consumption, since devices need to accumulate more activations in memory.
In Fig.~\ref{fig:safetystocks_cyclic}, since 7 micro-batches are injected at the beginning of the schedule (compared to 4 in 1F1B), the activation memory of maximum 7 micro-batches needs to be accumulated.
Conversely, we can reduce memory consumption by delaying micro-batch injection until previously accumulated activations are freed up by the backward pass.
In Fig.~\ref{fig:safetystocks_cyclic_memlimit}, we delay the injection of memory-consuming micro-batches 3 and 6 until backward pass of micro-batches 0 to 2 and 3 to 5 have been executed, freeing up more activation memory.
This reduces the peak accumulated activation to 3 micro-batches.

\paragraph{Memory-aware adaptive scheduling algorithm.}
To maximize throughput while limiting peak memory consumption, we dynamically decide the injection or delayed execution of micro-batches in the pipeline.
We give our memory-aware adaptive scheduling in Alg.~\ref{algo:pipeline_schedule}.
During execution scheduling, each device keeps track of the current memory consumption (lines 9, 15).
On scheduling a forward pass of a micro-batch, if memory consumption exceeds the device memory limit, the device will skip forward passes until backward passes have freed up enough memory to avoid OOM (line 14).
In this way, the training can continue without OOM as long as the activation of \textit{one} single micro-batch fits into device memory.

\begin{algorithm}[!t]
\caption{Memory-aware Adaptive Scheduling}\label{algo:pipeline_schedule}
\DontPrintSemicolon
\SetNoFillComment

\SetKwInOut{KwInputs}{Inputs}
\SetKwInOut{KwOutputs}{Outputs}

\KwInputs{$C$ - the number of devices (stages), $M$
- the number of micro-batches, $a_{i, j}$ - activation memory of micro-batch $i$ on the $j$th device; $l_j$ - memory limit of device $j$}
\KwOutputs{$O_j$ - micro-batch execution order on device $j$}

$O_j \gets [], \forall j\in[1, C]$\;
\tcc{Op buffers $S^f_j, S^b_j$, current memory $m_j$ for device $j$}
$S^f_j, S^b_j \gets [], m_j \gets 0, \forall j\in[1, C]$\;
\tcc{Initialize forward buffer on device 1}
$S^f_1 \gets [a_{1,1}, a_{2,1}, \ldots, a_{M,1}]$\; 
\While{$\exists j$ s.t. $S^f_j \neq []$ or $S^b_j \neq []$}{
    \tcc{New ops unlocked this cycle}
    $N^f_j, N^b_j \gets [], \forall j \in [1, C]$\;
	\ForEach{device $j$}{
        \If(\tcp*[f]{Schedule a backward op}){$S^b_j \neq []$}{
            $a_{i,j} = S^b_j$.pop(0) \tcp*{get first op in buffer}
            $m_j$ -= $a_{i, j}$ \tcp*{update memory}
            $O_j$ += $(i, \text{`B'})$ \tcp*{record op order}
            append next stage op to corresponding $N^b_{j^*}$, if exists
        }
        \If(\tcp*[f]{Schedule a forward op}){$S^f_j \neq []$}{
            $a_{i,j} = S^f_j$.pop(0) \tcp*{get first op in buffer}
            \If{$m_j + a_{i,j} < l_j$}{
                $m_j$ += $a_{i, j}$ \tcp*{update memory}
                $O_j$ += $(i, \text{`F'})$ \tcp*{record op order}
                append next stage op to the corresponding $N^f_{j^*}$ or $N^b_{j^*}$, if exists
            }
            \Else{
                $S^f_j$.prepend($a_{i,j}$)
            }
        }
	}
    $S^f_j \gets S^f_j + N^f_j, S^b_j \gets S^b_j + N^b_j$\;
    
}
\end{algorithm}

\paragraph{Micro-batch ordering.}
Our memory-aware adaptive schedule algorithm assumes an ordered list of micro-batches as input.
The injection order of micro-batches could also impact throughput due to variations in micro-batch execution time.
Owing to the complexity of the scheduling problem, modeling this effect, let alone optimizing it, is considerably challenging. 
We address this issue by clustering the micro-batches by predicted execution time (using our cost model), assuming that micro-batches with similar execution time should be scheduled in proximity, and permute execution order of the resulting clusters to find the best order attaining highest throughput.
We find that merely 3 or 4 clusters are adequate to attain satisfactory performance (i.e., throughput increase is insignificant when further increasing the number of clusters) through empirical study.

%% file: sections/async_comm_planning.tex
\section{Communication Planning}
\label{sec:comm_planning}

\begin{figure}[t]
    \centering
    \includegraphics[width=\linewidth]{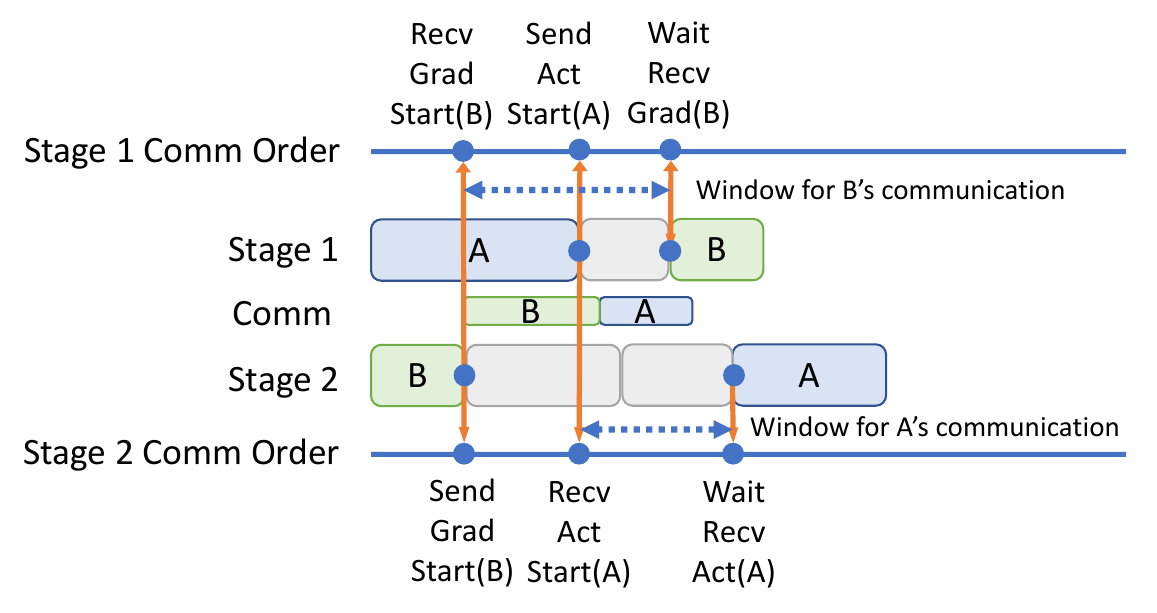}
    \caption{
    Planning the order of send and receive operations using simulated execution timeline. 
    }
    \label{fig:async_comm_planning}
\end{figure}

Given the micro-batches and their execution order in a training iteration, we further generate a communication plan specifying the order of all communication between pipeline stages (i.e., sending activation to the next stage and receiving gradient from the previous stage for each micro-batch at each stage).
To avoid deadlocking, we need to make sure that all pairs of sends and receives are executed in the same order on adjacent stages.
We guarantee this by scheduling \textit{both} send and receive operations together at the time where the tensor to be sent is generated.

Specifically, we simulate a device computation timeline using the pipeline schedule generated in Sec.~\ref{sec:pipeline_schedule} and the micro-batch execution time cost model, as shown in Fig.~\ref{fig:async_comm_planning}.
Then, we iterate through operations (ops) in the timeline (forward and backward of micro-batches) in ascending order of their end time, while maintaining a queue for each pipeline stage to record the communication order.
Upon processing each op, we push the corresponding send \textit{Start} op (\textit{SendActStart} for forward passes, and \textit{SendGradStart} for backward passes) into the queue of the executing stage.
At the same time, we also insert the matching receive \textit{Start} op into the receiver's queue.
The leftmost two vertical orange lines in Fig.~\ref{fig:async_comm_planning}, demonstrates this process.

To minimize blocking, we schedule the corresponding \textit{Wait} ops as late as possible, i.e. we only add \textit{Wait} ops before computations that consume the received tensor (the last two orange lines in Fig.~\ref{fig:async_comm_planning}).
This maximizes the time window in which communication can execute without blocking computation.

The pipeline execution and communication schedule is specified as a sequence of instructions for each device.
We then calculate the shapes of the communicated tensors based on the shape of input micro-batches and the model architecture, and include the shapes in the generated instructions to avoid exchanging them at runtime.

%% file: sections/other_optimizations.tex
\section{Implementation and Other Optimizations} 

\SystemName{} is implemented using 10K LoC in Python, with additional 500 LoC in C++ for accelerating the DP algorithm.
We use Redis~\cite{redis} as our distributed \textit{instruction store}.
Communication in pipeline training is implemented based on PyTorch's distributed communication package with NCCL~\cite{nvidia2023nccl} backend.
We implement the set of \textit{instructions} in around 400 LoC in Megatron-LM~\cite{narayanan2021efficientmlm} with PyTorch nightly version 2.1.0.dev20230322+cu117.
We further enable ZeRO~\cite{rajbhandari2020zero} optimizer by integrating Megatron-LM with DeepSpeed~\cite{deepspeed} version 0.9.1 since it's often used together with data parallelism.
We adopt the Megatron-LM's implementation for data and tensor parallelism while apply \SystemName{} to replace its pipeline modules, enabling training with mixed 3D parallelism.

To leverage \SystemName{} with Megatron-LM, users can directly reuse existing training scripts with additional arguments specifying configurations for \SystemName{} (e.g., the device memory limit and the number of CPU cores to use during planning).
To extend \SystemName{} to frameworks other than Megatron-LM, users only need to implement the \textit{pipeline instructions} in the framework.

\paragraph{Dynamic recomputation.}
Activation checkpointing (recomputation)~\cite{chen2016training} is a widely-used technique to reduce memory consumption during DNN training, by recomputing the activations during backward pass instead of storing them.
However, they come with extra computation cost.
There are also multiple ways to apply recomputation (e.g.,~\cite{narayanan2021efficientmlm}), resulting in different trade-offs between training throughput and memory consumption.
Under dynamic micro-batching, the peak memory consumption in different training iterations varies; thus we dynamically decide the best recomputation scheme for each iteration (i.e., the one with the least computation overhead without triggering OOM). 
This is achieved by repeating scheduling and micro-batch partitioning under different assumptions for recomputation method (using different cost-models).

\paragraph{Reducing memory fragmentation.}
Dynamic tensor shapes exacerbate the pressures to caching memory allocators (e.g., in PyTorch) since their dynamic memory requirement causes frequent cache misses.
Under memory pressure, we sometimes observe blocking \textit{cudaMalloc}'s and \textit{cudaFree}'s during training, which are caused by the allocator failing to find a usable memory block and PyTorch's effort to defragment GPU memory.
To reduce training slow-down caused by these blocking operations, we instruct PyTorch to use a single unified memory pool to manage all CUDA memory, and pre-allocate all GPU memory into the pool before training starts.
This eliminates the need for allocating (and freeing) CUDA memory during runtime.

%% file: sections/evaluation.tex
\section{Evaluation}
\label{sec:evaluation}

\paragraph{Testbed Set-up}
We conduct our experiments in a cluster of 4 Amazon EC2 p4d.24xlarge instances (32 GPUs in total).
Each p4d node is equipped with 8 NVIDIA A100 (40GB) GPUs and 96 vCPU cores.
NVSwitch connects the GPUs within each node, and the nodes are connected by a 400Gbps network with EFA~\cite{aws2022elastic} enabled.

\paragraph{DNNs}
We evaluate \SystemName{} by training two popular LLM models: GPT~\cite{brown2020gpt3} (decoder-only architecture) and T5~\cite{raffel2020t5} (encoder-decoder architecture).
We evaluate each model on four different cluster sizes, with the model size scaling accordingly.
For GPT, we scale the model parameters following the configuration in the GPT-3 paper~\cite{brown2020gpt3}.
For T5, since T5-11B is already the largest model specified in its paper and its hidden size in feed forward layers is huge, we simply scale the number of layers.
We list the detailed model specifications in Table~\ref{tab:model_specs}.

\begin{table*}[ht]
\small
\centering
\begin{tabular}{|c|c|c|c|c|c|c|c|}
\hline
Model & \# GPUs & \# layers & Model Dim & \# Heads & \# KV Channels & FFN Dim & \# Param (B) \\
\hline
GPT & 4,8,16,32 & 16,32,40,16 & 4096,4096,5140,12288 & 32,32,40,96 & 128 & 16384,16384,20560,49152 & 3.35,6.7,13,29 \\
\hline
T5 & 4,8,16,32 & 12,24,48,96 & 1024 & 128 & 128 & 65536 & 5.5,11,22,44 \\
\hline
\end{tabular}
\caption{DNN model configurations. For T5, ``\# layers'' refers to layers present in both the encoder and the decoder.}
\label{tab:model_specs}
\end{table*}

\paragraph{Dataset}
We use the zero-shot version of the FLANv2~\cite{Longpre2022flanv2} dataset in our experiments, which consists of 1836 different tasks and is one of the largest public multi-task training data collections.
The full dataset contains 15M training samples.
To reduce evaluation costs, we randomly down-sample it to 100K samples.
All our metrics reported are collected during one epoch of training on the down-sampled dataset.

\paragraph{Baselines}
We use Megatron-LM integrated with DeepSpeed (\textit{MLM+DS}) as the training system baseline, which implements packing (i.e., pack multiple sample into the same sequence so the resulting sequence length matches the specified maximum sequence length).
For each experiment, we grid search through common 3D parallelism combinations (power of twos in each of the data, tensor and pipeline parallel dimensions, with tensor parallelism limited to intra-node only) for both baselines and \SystemName{} to use on the given cluster configuration.
\SystemName{} implements 3D parallelism by reusing the data and tensor parallel implementation of Megatron-LM, but replacing its pipeline modules with our \textit{executor}. 
For baselines, we grid search common combinations of micro-batch size and activation checkpointing strategy, and report the best results.

\paragraph{Metrics}
We report the system throughput in terms of actual tokens processed, which does not count the padding tokens.
Specifically, throughput is calculated by dividing the total number of tokens in the training dataset by the amount of time needed for one epoch of training.

\subsection{Throughput under sequence length scaling}

\begin{figure*}[ht]
    \centering
     \begin{subfigure}[b]{0.245\linewidth}
     \centering
     \includegraphics[width=\linewidth]{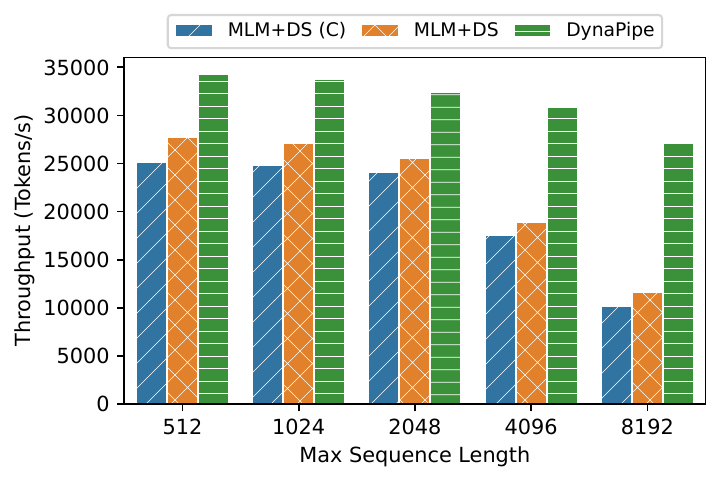}
     \caption{GPT (3.35B) on 4 GPUs.}
     \label{fig:throughput_gpt_small}
    \end{subfigure}
     \hfill
     \begin{subfigure}[b]{0.245\linewidth}
         \centering
         \includegraphics[width=\linewidth]{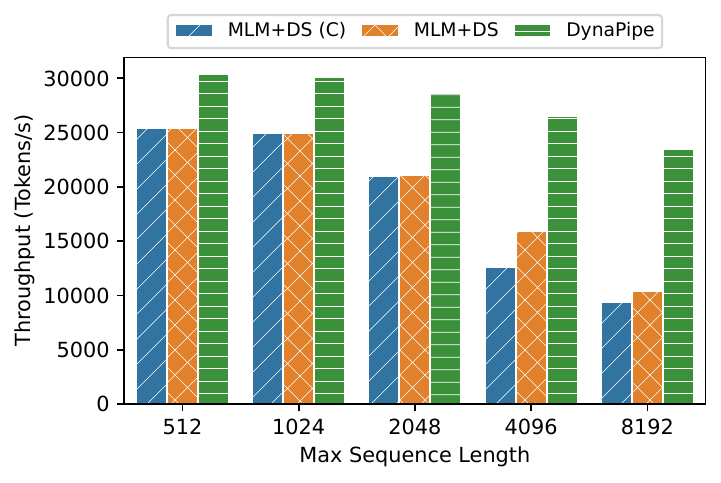}
         \caption{GPT (6.7B) on 8 GPUs.}
         \label{fig:throughput_gpt_mid}
     \end{subfigure}
      \hfill
     \begin{subfigure}[b]{0.245\linewidth}
         \centering
         \includegraphics[width=\linewidth]{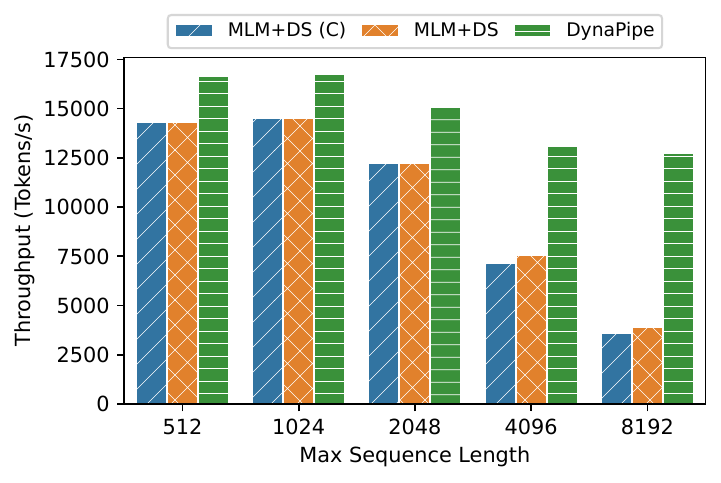}
         \caption{GPT (13B) on 16 GPUs.}
         \label{fig:throughput_gpt_large}
     \end{subfigure}
     \hfill
     \begin{subfigure}[b]{0.245\linewidth}
         \centering
         \includegraphics[width=\linewidth]{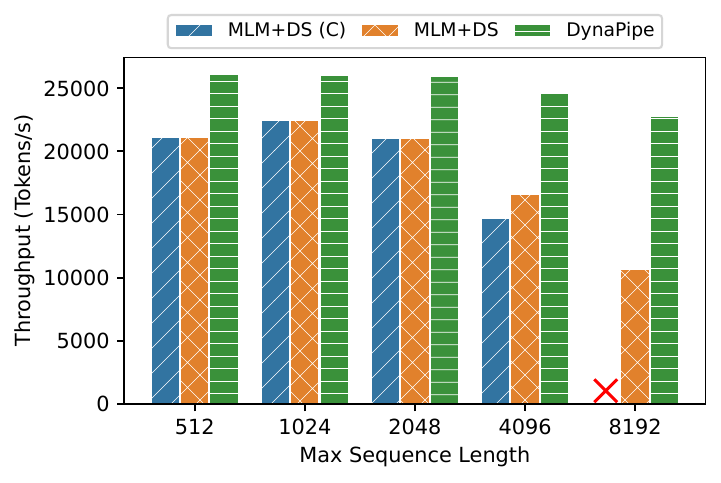}
         \caption{GPT (29B) on 32 GPUs.}
         \label{fig:throughput_gpt_xlarge}
     \end{subfigure}
      \hfill
     \begin{subfigure}[b]{0.245\linewidth}
         \centering
         \includegraphics[width=\linewidth]{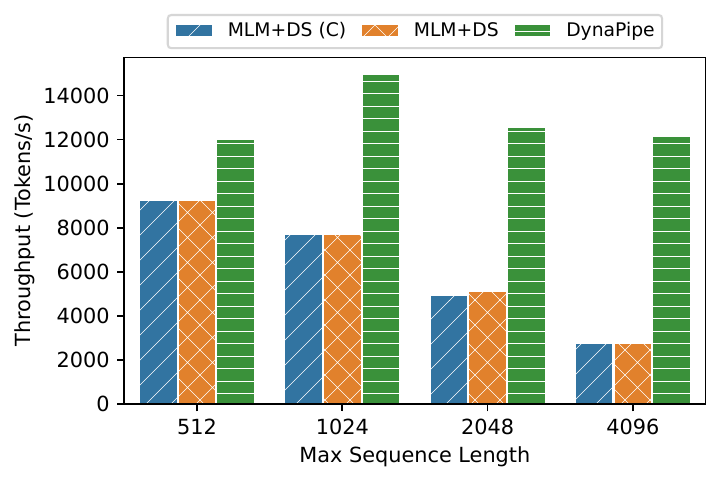}
         \caption{T5 (5.5B) on 4 GPUs.}
         \label{fig:throughput_t5_small}
     \end{subfigure}
       \hfill
     \begin{subfigure}[b]{0.245\linewidth}
         \centering
         \includegraphics[width=\linewidth]{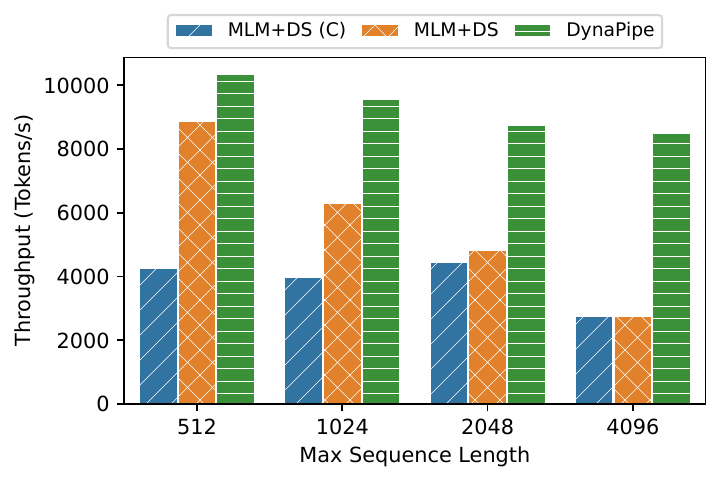}
         \caption{T5 (11B) on 8 GPUs.}
         \label{fig:throughput_t5_mid}
     \end{subfigure}
    \hfill
     \begin{subfigure}[b]{0.245\linewidth}
         \centering
         \includegraphics[width=\linewidth]{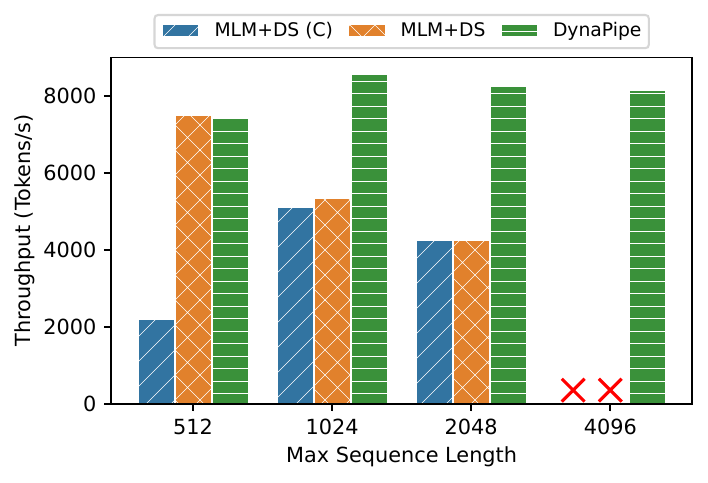}
         \caption{T5 (22B) on 16 GPUs.}
         \label{fig:throughput_t5_large}
     \end{subfigure}
     \hfill
     \begin{subfigure}[b]{0.245\linewidth}
         \centering
         \includegraphics[width=\linewidth]{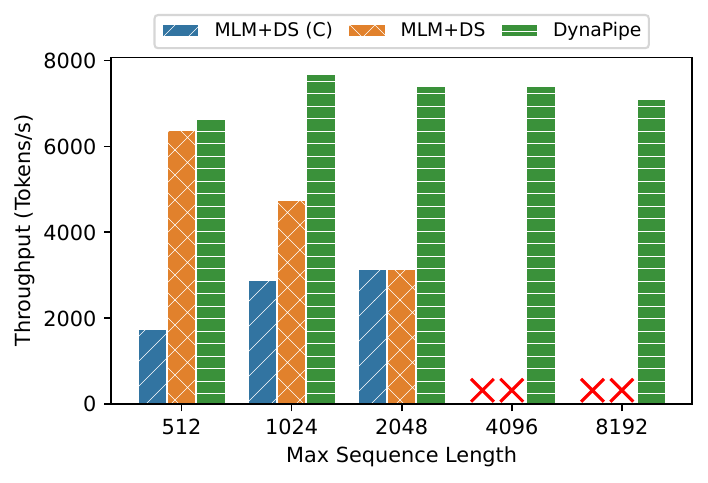}
         \caption{T5 (44B) on 32 GPUs.}
         \label{fig:throughput_t5_xlarge}
     \end{subfigure}
    \caption{Training throughput under different maximum sequence lengths.}
    \label{fig:throughput_scaling_seqlen}
\end{figure*}

\begin{figure*}[ht]
    \centering
     \begin{subfigure}[b]{0.245\linewidth}
     \centering
     \includegraphics[width=\linewidth]{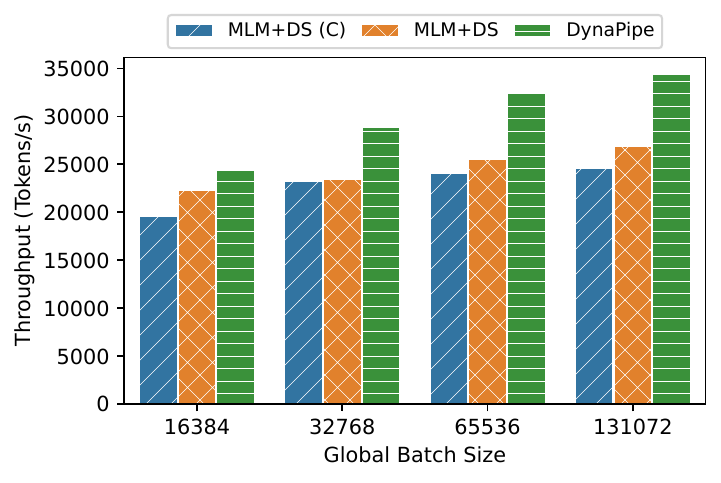}
     \caption{GPT (3.35B) on 4 GPUs.}
     \label{fig:throughput_gpt_small_gbs}
    \end{subfigure}
     \hfill
     \begin{subfigure}[b]{0.245\linewidth}
         \centering
         \includegraphics[width=\linewidth]{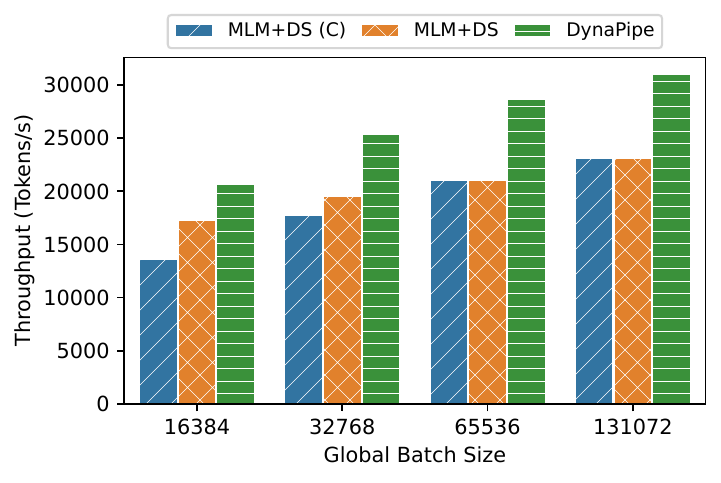}
         \caption{GPT (6.7B) on 8 GPUs.}
         \label{fig:throughput_gpt_mid_gbs}
     \end{subfigure}
      \hfill
     \begin{subfigure}[b]{0.245\linewidth}
         \centering
         \includegraphics[width=\linewidth]{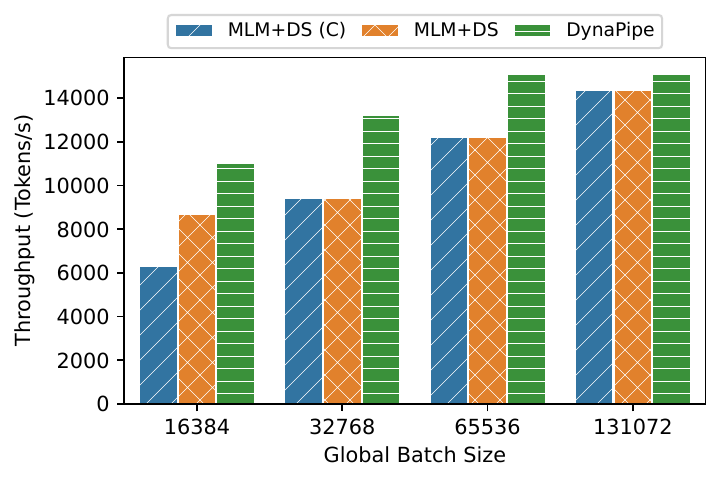}
         \caption{GPT (13B) on 16 GPUs.}
         \label{fig:throughput_gpt_large_gbs}
     \end{subfigure}
     \hfill
     \begin{subfigure}[b]{0.245\linewidth}
         \centering
         \includegraphics[width=\linewidth]{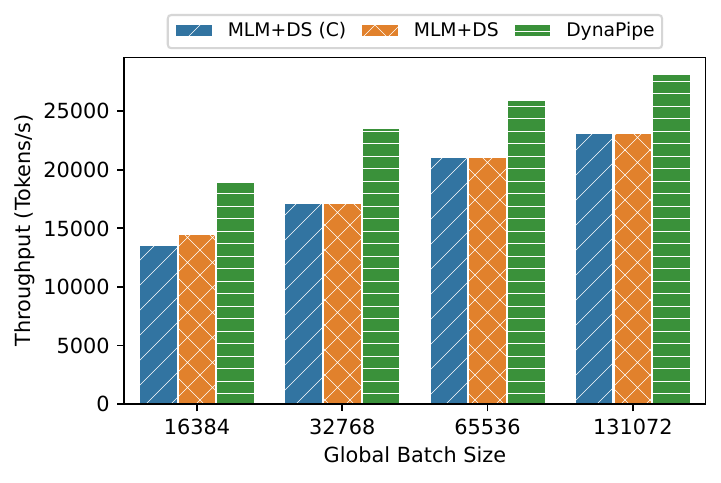}
         \caption{GPT (29B) on 32 GPUs.}
         \label{fig:throughput_gpt_xlarge_gbs}
     \end{subfigure}
      \hfill
     \begin{subfigure}[b]{0.245\linewidth}
         \centering
         \includegraphics[width=\linewidth]{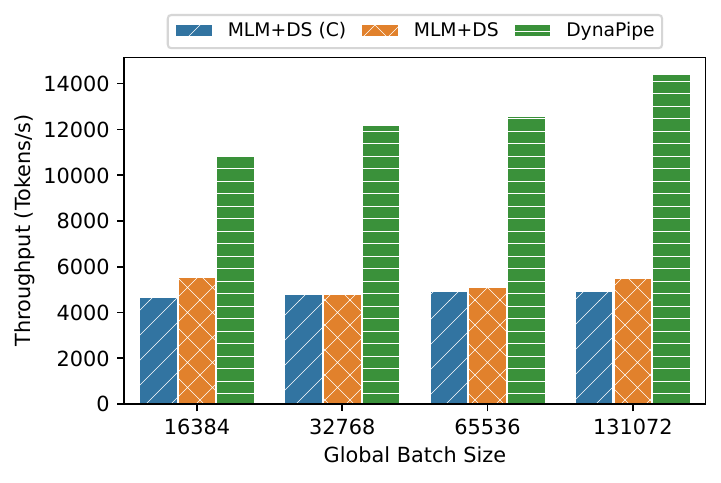}
         \caption{T5 (5.5B) on 4 GPUs.}
         \label{fig:throughput_t5_small_gbs}
     \end{subfigure}
       \hfill
     \begin{subfigure}[b]{0.245\linewidth}
         \centering
         \includegraphics[width=\linewidth]{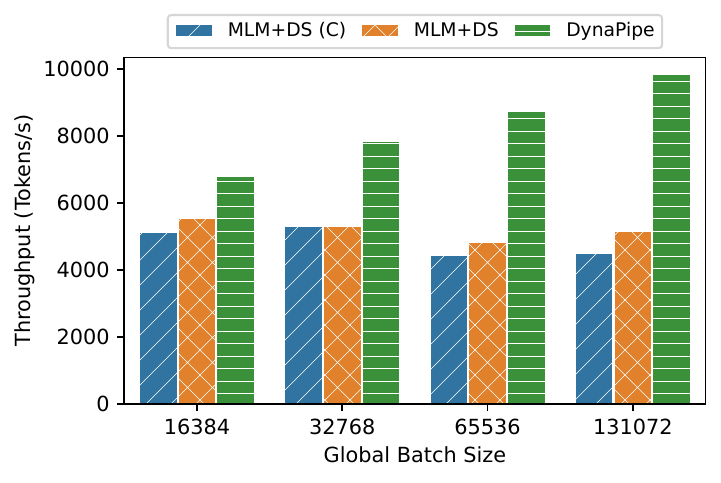}
         \caption{T5 (11B) on 8 GPUs.}
         \label{fig:throughput_t5_mid_gbs}
     \end{subfigure}
    \hfill
     \begin{subfigure}[b]{0.245\linewidth}
         \centering
         \includegraphics[width=\linewidth]{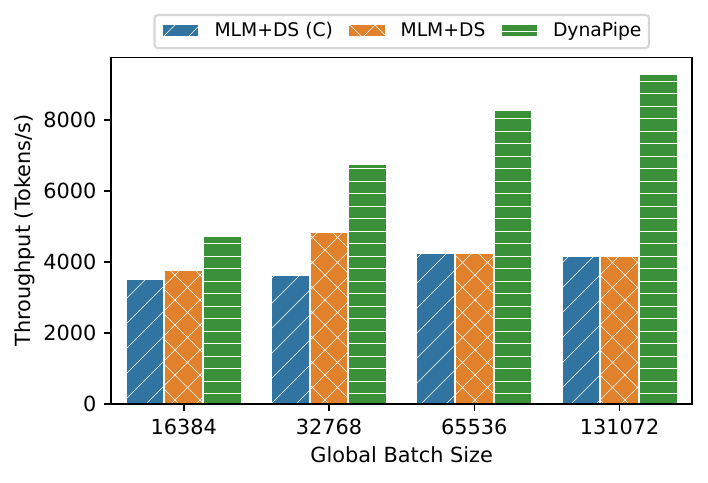}
         \caption{T5 (22B) on 16 GPUs.}
         \label{fig:throughput_t5_large_gbs}
     \end{subfigure}
     \hfill
     \begin{subfigure}[b]{0.245\linewidth}
         \centering
         \includegraphics[width=\linewidth]{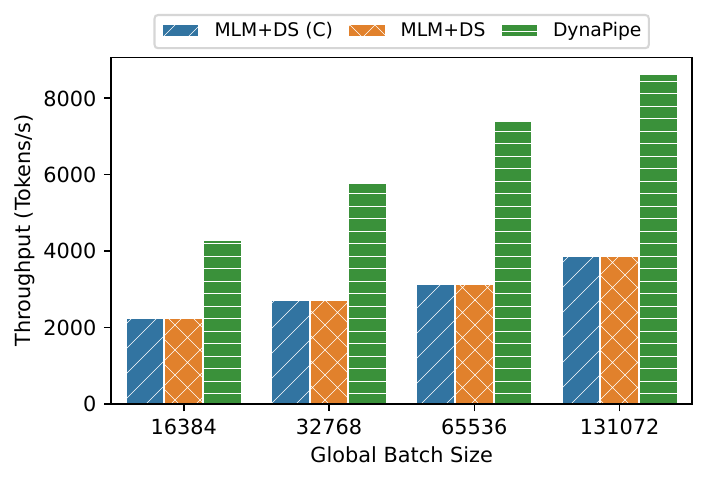}
         \caption{T5 (44B) on 32 GPUs.}
         \label{fig:throughput_t5_xlarge_gbs}
     \end{subfigure}
    \caption{Training throughput under different global batch sizes.}
    \label{fig:throughput_scaling_gbs}
\end{figure*}

We fix the global batch size (i.e., size of the input mini-batch in each training iteration, to be divided among devices if data parallelism is used) to 65536 tokens and vary the maximum sequence length allowed in the dataset (sequences that are longer are truncated).
Since the baseline (\textit{MLM+DS}) and \SystemName{} may achieve maximum throughput under different grid-searched parallelisms, we also evaluate the throughput of the baseline when it uses the same parallelism configuration as \SystemName{} (\textit{MLM+DS (c)}).

In Fig.~\ref{fig:throughput_scaling_seqlen}, we observe that in most cases, the throughput of \textit{MLM+DS} decreases rapidly as maximum sequence length scales up, due to the super-linear relationship between computation time and maximum sequence length (Fig.~\ref{fig:time_vs_seqlen}).
\SystemName{} dynamically decides the sequence lengths in micro-batches, and its performance is determined more by the average sequence length than the maximum.
As a result, while we still see throughput decrease in \SystemName{} as maximum sequence length increases (since we allow longer sequence to exist without truncation), the decrease is much less than that of \textit{MLM+DS}.

On T5 (16 and 32 GPUS), \SystemName{} scales to higher sequence lengths than baselines.
This is because our memory-aware adaptive schedule can dynamically adjust the schedule to accommodate memory limit (Fig.~\ref{fig:safetystocks_cyclic_memlimit}), achieving a lower peak memory consumption than 1F1B which is used by baselines.
In many cases (e.g. T5 on 8 GPUs), we also find the MLM+DS under-performs using the optimal parallelism for \SystemName{}.
This indicates that the performance characteristic is very different for \SystemName{} and packing-based baselines.

\subsection{Throughput under global batch size scaling}

In Fig.~\ref{fig:throughput_scaling_gbs}, we set the maximum sequence length to 2048 and adjust the global batch size.
For the GPT model, both MLM+DS and \SystemName{}'s performance increases with global batch size, since larger global batch sizes reduce the synchronization frequency of data parallelism and the size of the pipeline bubble in pipeline parallelism.
We also observe the performance of \SystemName{} increase faster than MLM+DS, since \SystemName{} also benefits from the increased opportunity to optimize micro-batch splitting when global batch size is large.
For T5 model, the baselines make and extensive use of tensor parallelism whose performance is less affected by the global batch size.
While \SystemName{} uses higher degree of pipeline parallelism, therefore still see the performance improvement when global batch size scales up.

\subsection{Padding efficiency}

\begin{figure}[ht]
    \centering
     \begin{subfigure}[b]{\columnwidth}
     \centering
     \includegraphics[width=0.49\linewidth]{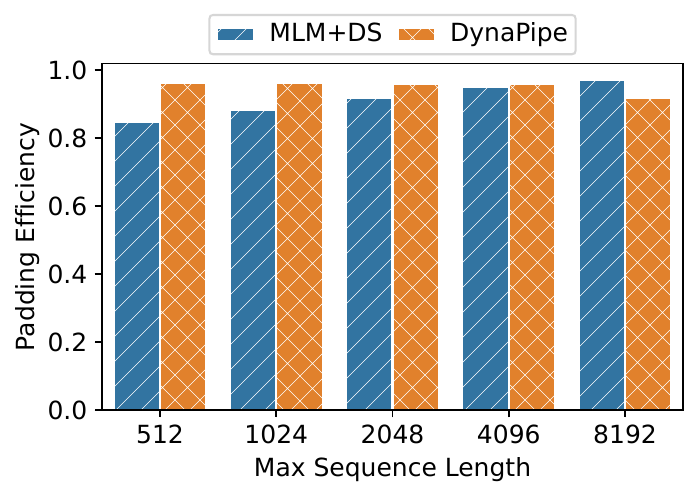}
     \hfill
      \includegraphics[width=0.49\linewidth]{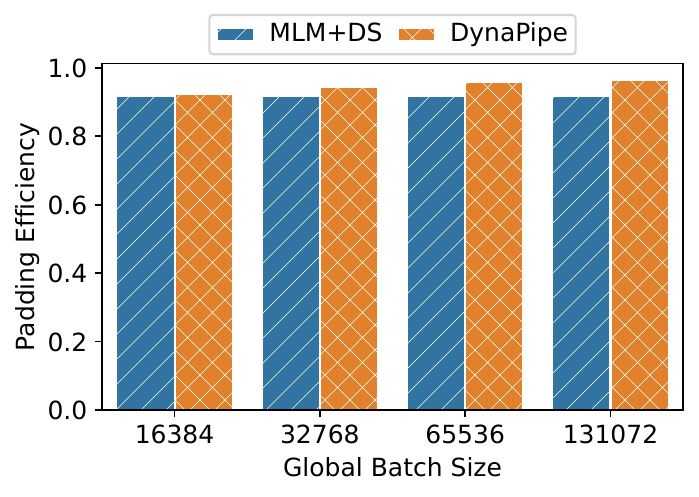}
     \caption{GPT (6.7B) on 8 GPUs}
     \label{fig:padding_eff_gpt}
    \end{subfigure}
     \hfill
     \begin{subfigure}[b]{\columnwidth}
         \centering
         \includegraphics[width=0.49\linewidth]{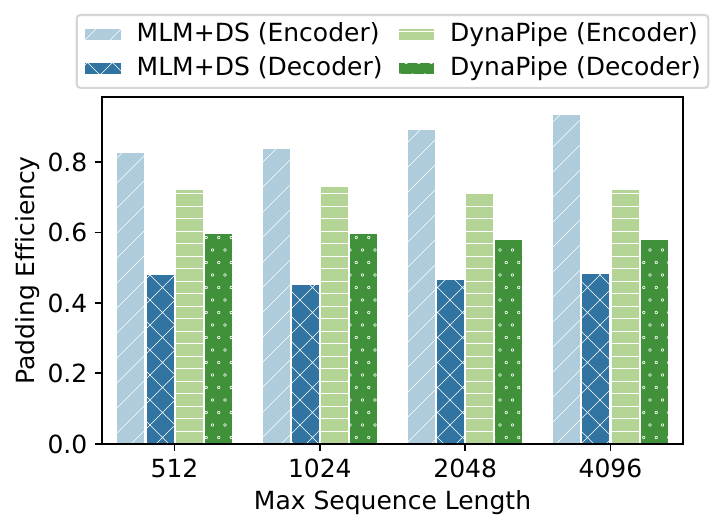}
         \hfill
          \includegraphics[width=0.49\linewidth]{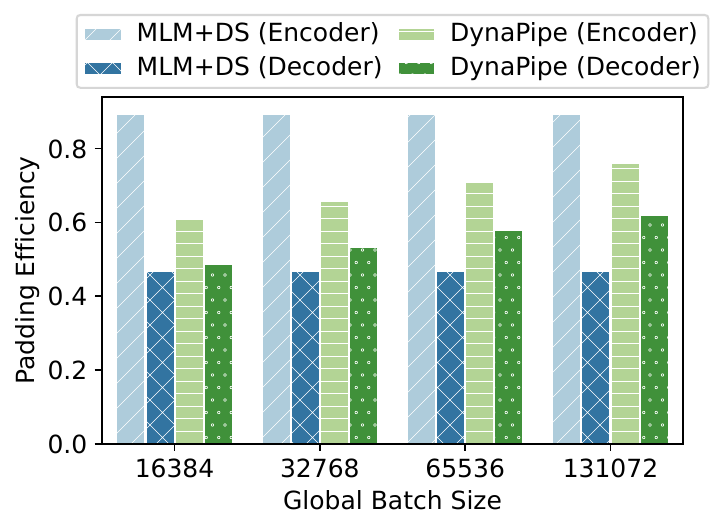}
         \caption{T5 (11B) on 8 GPUs}
         \label{fig:padding_eff_t5_seqlen}
     \end{subfigure}
    \caption{Padding efficiency case study.
    }
    \label{fig:batching_efficiency}
\end{figure}

For GPT models, both packing and our dynamic micro-batching can achieve a high padding efficiency (>0.8, Fig.~\ref{fig:padding_eff_gpt}), with ours slightly higher. However, as we can see from the throughput results in Fig.~\ref{fig:throughput_gpt_small}, high padding efficiency does not directly translate to better throughput.
In Fig.~\ref{fig:padding_eff_gpt}, we also see the padding efficiency of packing increases with maximum sequence length.
For T5 models, packing has a very high padding efficiency in terms of input to the encoder, while padding efficiency to the decoder is much lower.
Our padding efficiency is more balanced between encoder and decoder, since we consider both input sequence lengths during our DP algorithm.

\subsection{Ablation study}
\label{sec:ablation}

\begin{figure}[t]
    \centering
     \begin{subfigure}[b]{0.49\columnwidth}
     \centering
     \includegraphics[width=0.9\linewidth]{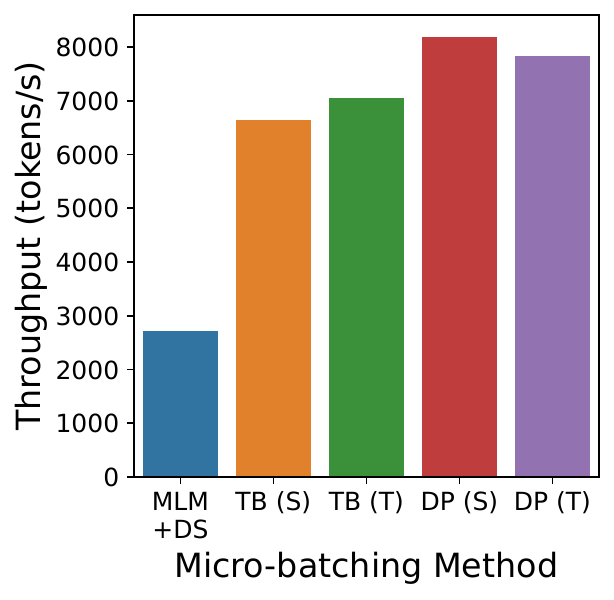}
     \caption{Micro-batching methods.}
     \label{fig:ablation_batching}
    \end{subfigure}
      \hfill
       \begin{subfigure}[b]{0.49\columnwidth}
         \centering
         \includegraphics[width=\linewidth]{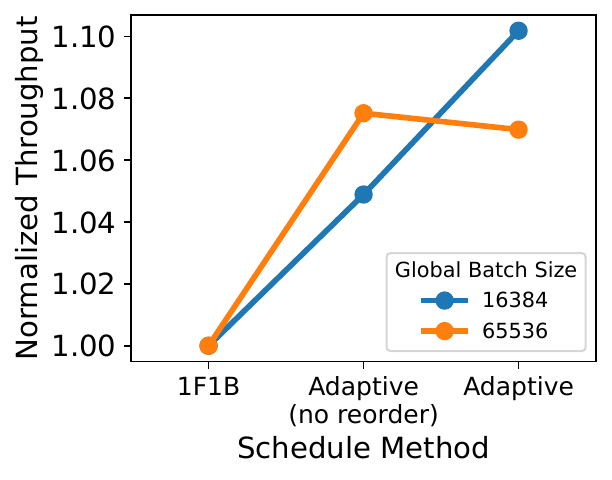}
         \caption{Pipeline schedule methods.}
         \label{fig:ablation_schedule}
     \end{subfigure}
    \caption{Ablation study.}
    \label{fig:ablation_study}
\end{figure}

We assess the design components within \SystemName{} to analyze and decompose its performance improvement.

We first compare our dynamic programming algorithm against packing in \textit{MLM+DS} and token-based (\textit{TB}) micro-batching (which splits micro-batches so that each micro-batch contains roughly the same number of tokens), when training T5 with maximum sequence length 4096 and global batch size 65536 on 8 GPUs in Fig.~\ref{fig:ablation_batching}.
In this setting, the optimal parallelism configuration does not use pipelining, isolating the impact of micro-batching.
After searching for the best number of tokens per micro-batch, we find that \textit{TB} already achieves significantly higher throughput than \textit{MLM+DS}, indicating the inefficiency of packing-based solutions.
Our dynamic programming algorithm further out-performs \textit{TB} (without the need for parameter searching) by striking better trade-offs among padding, computation efficiency and memory consumption.
We also evaluate different ways of determining the order of samples in our dynamic micro-batching and the TB baseline, i.e., sorting (S) versus solving a traveling salesman problem using a solver (T), and find that they do not impact the performance much.

We next compare our pipeline schedule (adaptive scheduling with and without micro-batch ordering) with 1F1B schedule, when training GPT with the same maximum sequence length and GPU settings as above.
The grid-searched best parallelism uses 4 pipeline stages.
As shown in Fig.~\ref{fig:ablation_schedule}, our pipeline schedule achieves 10.1\% and 7.4\% throughput improvement over 1F1B, under global batch size 16384 and 65536, respectively.
The performance gain is less than that from our simulations (Fig.~\ref{fig:schedule_under_variation}), since now our dynamic micro-batching includes the longest micro-batch execution time in the cost function (\ref{eqn:t_iter}), which, when minimized, produces more uniform micro-batches.
The effect of micro-batch ordering is less prominent under a large global batch size, since sequence lengths in micro-batches can be made more similar by our DP algorithm when the global batch size is large (due to more opportunities for better micro-batch splitting).

\subsection{Execution planning time}

\begin{figure}[t]
    \centering
     \begin{subfigure}[b]{\columnwidth}
     \centering
     \includegraphics[width=0.7\linewidth]{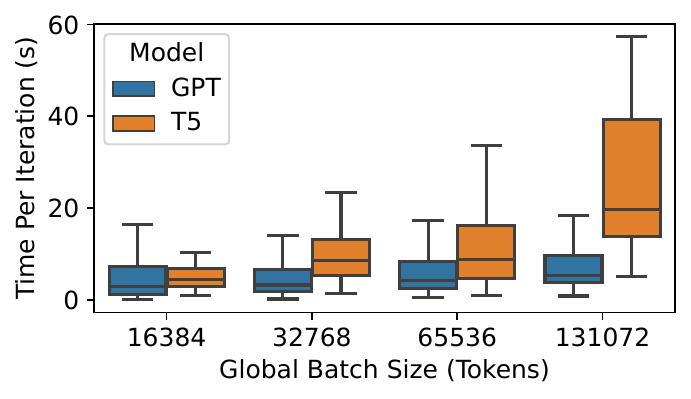}
     \caption{Planning time distribution.}
     \label{fig:planning_time_distribution}
    \end{subfigure}
     \begin{subfigure}[b]{\columnwidth}
     \centering
     \includegraphics[width=0.7\linewidth]{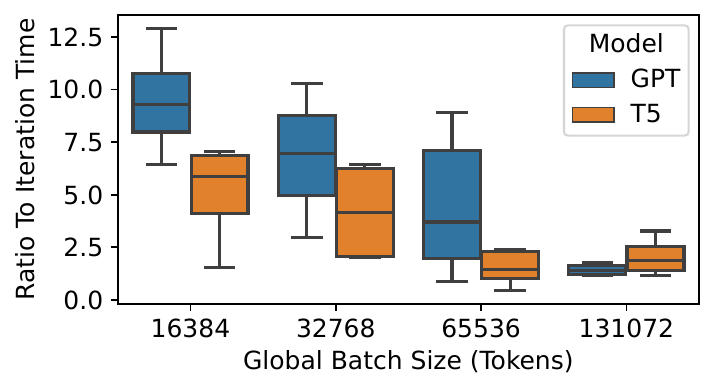}
     \caption{Comparison between average planning time and average iteration time (planning time / iteration time).}
     \label{fig:planning_time_comparison}
    \end{subfigure}
    \caption{
    Execution planning time.
    }
    \label{fig:preprocessing_time}
\end{figure}

We present the single-thread execution plan generation time during all our experiments in Fig.~\ref{fig:planning_time_distribution}.
Execution planning for most training iterations takes less than 20 seconds for both GPT and T5 when the global batch size is small.
Under larger global batch sizes, the time needed for our dynamic micro-batch and scheduling algorithms increases.
The execution planning can be parallelized among the large number of CPU cores or even multiple nodes, allowing it to be completely overlapped with GPU computation.
Fig.~\ref{fig:planning_time_comparison} compares our planning time to actual execution time per iteration.
Across all our experiments, the average planning time to iteration time ratio peaks at 12.9x, which means full overlapping between training and planning is feasible with only 13 CPU cores, much less than available cores in typical LLM training instances (Amazon EC2 p4d~\cite{AmazonEC2p4d}, Microsoft Azure ND A100 v4~\cite{MSNDA100} and Google Could a2-highgpu-1g~\cite{GCPa2highgpu1g}, each possessing 96 vCPU cores).
In our experiments, we parallelize plan generation on 64 CPU cores in each machine that participates in training, and observe no slow-down caused by execution plan generation.

\subsection{Accuracy of cost models}
\label{sec:cost_model_accuracy}

\begin{figure}[t]
    \centering
     \begin{subfigure}[t]{0.49\columnwidth}
     \centering
     \includegraphics[width=\linewidth]{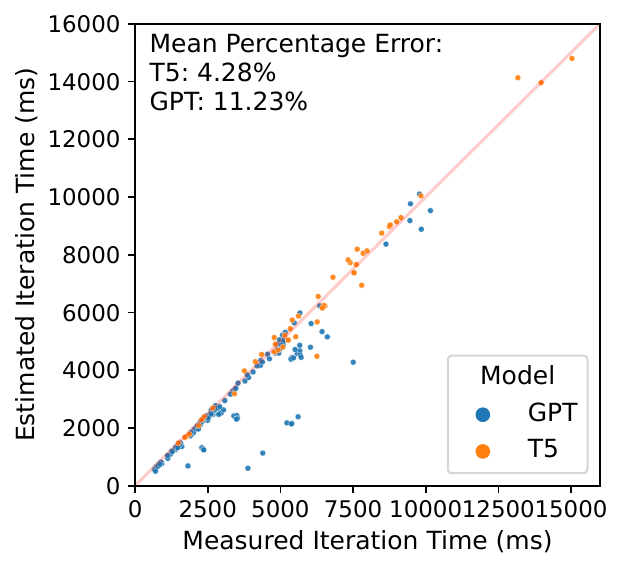}
     \caption{Iteration time.}
     \label{fig:cost_model_time}
    \end{subfigure}
      \hfill
     \begin{subfigure}[t]{0.49\columnwidth}
         \centering
         \includegraphics[width=\linewidth]{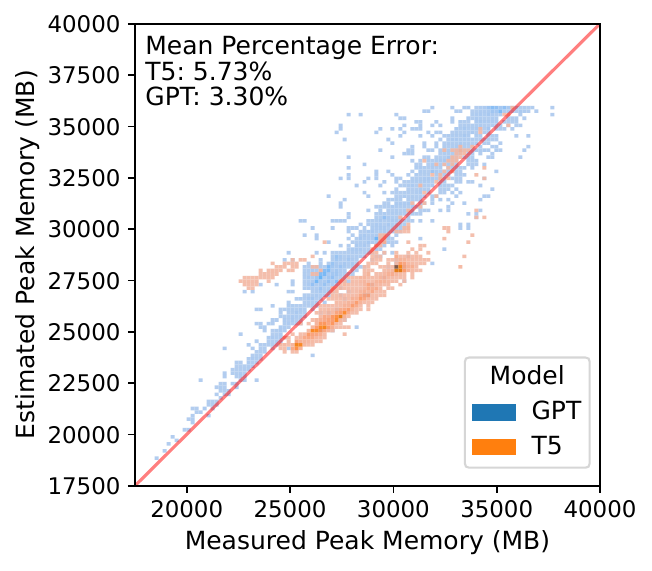}
         \caption{Peak Memory Consumption.}
         \label{fig:cost_model_memory}
     \end{subfigure}
    \caption{Prediction accuracy of our iteration time and peak memory cost models.}
    \label{fig:cost_model_accuracy}
\end{figure}

Fig.~\ref{fig:cost_model_accuracy} illustrates the prediction accuracy of our iteration time and memory cost models, where data points are collected from all our experiments.
For T5, the estimated iteration time matches well the actual value.
The iteration time prediction error for GPT is slightly larger, and the out-liners are due to the all-reduce operation in data parallelism it uses, which we do not model.
For memory estimation, we achieve lower than 6\% mean prediction error for both T5 and GPT.
These confirm that our cost models can provide accurate signals for guiding our optimizations.

\section{Related Works}

\paragraph{3D parallel training frameworks.}
Megatron-LM~\cite{narayanan2021efficientmlm}, DeepSpeed~\cite{deepspeed} are two popular frameworks supporting 3D parallel LLM training. 
Alpa~\cite{zheng2022alpa} further automates the parallelization of the model, considering both intra- (including but not limited to data and tensor parallelism) and inter-operator (i.e., pipeline) parallelism.
These frameworks assume fixed micro-batch sizes, fixed number of micro-batches and use a static pipeline schedule (e.g., 1F1B).
They also do not consider construction of micro-batches, leaving the decision to users.
\SystemName{} optimizes micro-batch construction, supports dynamic micro-batches, and adopts adaptive pipeline schedule to accelerate multi-task training.
Approaches that shard the model and optimizer states (e.g., ZeRO~\cite{rajbhandari2020zero}, FSDP~\cite{zhao2023fsdp}), or optimize communication in data parallelism (e.g., MiCS~\cite{zhang2022mics}) are also often used together with 3D parallelism.
These approaches are orthogonal to \SystemName{} and may be used in conjunction.

\paragraph{Sort dataset before batching.}
Some libraries (e.g., fairseq~\cite{ott2019fairseq} and tensor2tensor~\cite{tensor2tensor}) offer an option to sort the dataset before constructing the mini-batches, so each mini-batch will contain samples with similar sequence lengths (also referred to as bucketing).
Such bucketing destroys the randomness in batch construction thus may affect model performance.
\SystemName{} fully respects users' mini-batch construction method and only reorders samples within each mini-batch, preserving mathematical equivalence with models trained using traditional methods (padding or packing).

\paragraph{Custom attention kernels that ignore padding.}
ByteTransformer~\cite{zhai2023bytetransformer} implements special CUDA kernels to skip padding during self-attention. 
FlashAttention~\cite{dao2022flashattention} also include attention kernels allowing variable sequence lengths.
However, to ignore padding in other parts of the models, the model code needs to be adapted accordingly.
Both libraries also only offer padding-free implementation for the BERT~\cite{devlin2019bert} model.
Through dynamic micro-batching, \SystemName{} directly minimizes padding in the inputs without needing to modify any model components.
Recent updates (since the submission of \SystemName{}) in FlashAttention also includes a method to ``unpack'' samples from packed inputs, enabling the variable length attention kernels to be used in conjunction with packing.
It would be interesting future work to benchmark the performance of packing while enabling such kernel optimizations.
None of the above works discussed the performance implications when jointly used with data or pipeline parallelism.
\SystemName{} optimizes for hybrid 3D parallel training by balancing the data parallel model replicas during micro-batch construction and using efficient pipeline schedules for dynamic micro-batches.

\paragraph{Training LLMs with extremely long sequences.}
Algorithmic approaches like sparse attention~\cite{child2019generating} and Longformer~\cite{beltagy2020longformer} tries to lower the quadratic complexity of self-attention in sequence length.
Systematic approaches like DeepSpeed-Ulysses~\cite{jacobs2023dsulysses} and LightSeq~\cite{li2023lightseq} partition the model inputs at sequence dimension and distribute the calculation of self-attention to multiple machines.
\SystemName{} pursue different goals from these works since we do not aim to improve the computation speed or memory consumption for extremely long sequence lengths.
Instead, we tackle the problem of input sequence length variation and avoid packing short input samples to long sequences using dynamic micro-batching.
It would be interesting future work to study how to concurrently utilize these methods and \SystemName{} to achieve both goals.

\section{Conclusion}
In this paper, we propose {\SystemName{}}, which uses dynamic micro-batching as a solution for the highly variable sequence length in multi-task training, circumventing packing which is computationally inefficient.
We propose a dynamic programming algorithm to optimize micro-batch construction, present a pipeline scheduling algorithm that is robust to micro-batch execution time variations, and design an effective pipeline communication planning mechanism for efficient dynamic pipeline training. 
Extensive evaluation demonstrates up to 4.39x speed up when training T5, and 3.25x when training GPT.

\section{Acknowledgements}
We would like to thank the anonymous reviewers and our shepherd Jongsoo Park for their valuable feedback.
This work was supported by an Amazon Research Award (ARA) on AWS AI and grants from Hong Kong RGC under the contracts HKU 17208920 and C7004-22G (CRF).

%% file: sections/artifact_appendix.tex
%
\appendix
\section{Artifact Appendix} 

\subsection{Abstract}
This artifact appendix documents the steps to reproduce Fig. 13 through 18 in the paper \textit{DynaPipe: Optimizing Multi-task Training through Dynamic Pipelines}. The experiments are expected to be run on a single AWS EC2 p4d instance. The evaluation is expected to take around 26 hours to complete.

\subsection{Description \& Requirements}

\subsubsection{How to access}
\hfill\\
The code used to generate results in the paper is mainly implemented in two repositories:
\begin{itemize}
    \item DynaPipe: contains the main implementation of DynaPipe. Can be accessed through \url{https://github.com/awslabs/optimizing-multitask-training-through-dynamic-pipelines}
    \item A modified version of Megatron-LM: Main modifications include adding support for packing in the dataloader, implementing the pipeline instructions for DynaPipe, and adding the scripts for running the experiments. Can be accessed through \url{https://github.com/chenyu-jiang/Megatron-LM}
\end{itemize}
A permanent copy of the artifact can be found at \url{https://zenodo.org/records/8413925} (DOI: 10.5281/zenodo.8413925), which contains a copy of this document, a pre-built docker image containing all codes and dependency required to run the artifact, and pre-processed datasets used in the experiments.

\subsubsection{Hardware dependencies}
\hfill\\
The evaluation is expected to be performed on a single Amazon EC2 p4d instance. Evaluation on other platforms with multiple GPUs (which supports PyTorch and Megatron-LM) is also possible when generating everything from scratch (see later sections for details).

\subsubsection{Software dependencies}

\begin{itemize}
    \item PyTorch (>= 2.1.0)
    \item DynaPipe (please see \textit{requirements.txt} in the repository for details)
    \item Modified Megatron-LM (dependencies of original Megatron-LM still apply)
    \item Modified DeepSpeed: \url{https://github.com/chenyu-jiang/DeepSpeed}. We removed a timer that introduce unnecessary synchronization which disrupts our schedule and disabled overflow checking for more consistent throughput measurement.
\end{itemize}

\subsubsection{Benchmarks}
\hfill\\
We used the FLANv2 dataset in our experiments. For artifact evaluation, we provide a pre-downloaded and pre-processed dataset in the provided machine (also accessible in the Zenodo repository). To copy the pre-downloaded dataset into the container, run the following command outside of the container:
\begin{verbatim}
cd ~/preprocessed_datasets
docker cp datasets dynapipe:/root/Megatron-LM
\end{verbatim}

\noindent
Otherwise, the dataset can be downloaded using the following steps:

\begin{enumerate}
\item Clone the repository for the dataset (a fork of the original repository with some version mismatch fixed. Also added a downloading script) and install dependencies:
\begin{verbatim}
git clone https://github.com/chenyu-jiang/
text-to-text-transfer-transformer.git
cd text-to-text-transfer-transformer
pip3 install -r requirements.txt 
\end{verbatim}

\item Download the raw dataset 

(generates \textit{supervised\_proportional.jsonl}):
\begin{verbatim}
python3 prepare_dataset.py
\end{verbatim}

\item Perform some initial cleaning

(generates \textit{cleaned\_supervised\_proportional.jsonl}):
\begin{verbatim}
python3 clean_dataset.py
\end{verbatim}

\item Preprocess the dataset with Megatron-LM's data loader script (generates \textit{.bin} and \textit{.idx} files)
\begin{verbatim}
cd <path_to_modified_MegatronLM>
./experiment_scripts/run_preprocess_flan.sh 
<path_to_cleaned_jsonl>
\end{verbatim}
\end{enumerate}

\subsection{Set-up}

We provide a Dockerfile to setup a container image for evaluation. To generate the image, run:
\begin{verbatim}
git clone
  https://github.com/chenyu-jiang/Megatron-LM.git
cd Megatron-LM/docker
./build_image.sh
\end{verbatim}
For artifact evaluation, a pre-built image will be installed on the provided machine (also available in the Zenodo repository).
\noindent
To create a container from the image, run (inside the docker directory):
\begin{verbatim}
./run.sh
\end{verbatim}
You will find DynaPipe and the modified Megatron-LM at \textit{/root} in the container. 

\subsection{Evaluation workflow}

\subsubsection{Major Claims}

\begin{itemize}
    \item (C1): DynaPipe significantly outperforms the state-of-the-art packing-based systems for multi-task training. This is proven by the experiment (E1) described in Section 8 whose results are illustrated in Figure 13 and 14.
    \item (C2): DynaPipe achieves comparable batching efficiency as packing-based systems. The batching efficiency statistics are illustrated in Figure 15.
    \item (C3) The dynamic-programming-based micro-batch generation and memory-aware adaptive scheduling algorithms proposed in DynaPipe out-performs naive alternatives. This is proven by experiment (E2) in Section 8, results illustrated in Figure 16.
    \item (C4) The planning overhead of DynaPipe is low (when parallelized onto multiple CPU cores). The planning time distribution is illustrated in Figure 17.
    \item (C5) DynaPipe's cost models can achieve a high prediction accuracy. Cost model accuracy statistics is illustrated in Figure 18.
\end{itemize}

\subsubsection{Experiments}
\hfill\\
Note: we provide some pre-computed results for some very time-consuming steps like grid searching for the best parallelism. To produce everything from scratch, see the README in \url{ https://github.com/chenyu-jiang/Megatron-LM}. We also provide a single script for running all needed experiments at \textit{/root/Megatron-LM/experiment\_scripts/run\_all.sh}.

\hfill\\
\noindent
\textbf{Experiment (E1)}: [Throughput Benchmark] [18 compute-hours]: Uses pre-generated configs (obtained by the grid search) to run full benchmarks for throughput comparison.
Note for artifact evaluation, only Fig.13 (a)(b)(e)(f) and Fig.14 (a)(b)(e)(f) can be generated on a single p4d node.
The other sub-figures of Fig.13 and 14 require multiple p4d nodes.
For Fig.17, only Fig.17 (a) will be generated.
Verifies (C1),(C2),(C4),(C5).

\noindent
\textit{[How to]}
Run the following command in the docker container:
\begin{verbatim}
cd /root/Megatron-LM
# performs benchmark
# (generates raw results in ./experiments directory)
./experiment_scripts/run_benchmark.sh
# generate figures
./experiment_scripts/generate_figure_13_14.sh
./experiment_scripts/generate_figure_15.sh
./experiment_scripts/generate_figure_17.sh
./experiment_scripts/generate_figure_18.sh
\end{verbatim}
\textit{[Results]}
Figure 13, 14, 15, 17, 18 will be generated in \textit{/root/Megatron-LM/reproduced\_figures}.

\hfill\\
\noindent
\textbf{Experiment (E2)}: [Ablation] [8 compute-hours]: Performs ablation study to verify (C3).

\noindent
\textit{[How to]}
Run the following command in the docker container:
\begin{verbatim}
cd /root/Megatron-LM
# performs ablation experiments
# (generates raw results in ./experiments directory)
./experiment_scripts/run_ablation.sh
# generate figures
./experiment_scripts/generate_figure_16.sh
\end{verbatim}
\textit{[Results]}
Figure 16 will be generated in 

\noindent
\textit{/root/Megatron-LM/reproduced\_figures}.

%% file: main.bbl

\begin{thebibliography}{40}


\ifx \showCODEN    \undefined \def \showCODEN     #1{\unskip}     \fi
\ifx \showDOI      \undefined \def \showDOI       #1{#1}\fi
\ifx \showISBNx    \undefined \def \showISBNx     #1{\unskip}     \fi
\ifx \showISBNxiii \undefined \def \showISBNxiii  #1{\unskip}     \fi
\ifx \showISSN     \undefined \def \showISSN      #1{\unskip}     \fi
\ifx \showLCCN     \undefined \def \showLCCN      #1{\unskip}     \fi
\ifx \shownote     \undefined \def \shownote      #1{#1}          \fi
\ifx \showarticletitle \undefined \def \showarticletitle #1{#1}   \fi
\ifx \showURL      \undefined \def \showURL       {\relax}        \fi
\providecommand\bibfield[2]{#2}
\providecommand\bibinfo[2]{#2}
\providecommand\natexlab[1]{#1}
\providecommand\showeprint[2][]{arXiv:#2}

\bibitem[Aghajanyan et~al\mbox{.}(2021)]%
        {aghajanyan2021muppet}
\bibfield{author}{\bibinfo{person}{Armen Aghajanyan}, \bibinfo{person}{Anchit
  Gupta}, \bibinfo{person}{Akshat Shrivastava}, \bibinfo{person}{Xilun Chen},
  \bibinfo{person}{Luke Zettlemoyer}, {and} \bibinfo{person}{Sonal Gupta}.}
  \bibinfo{year}{2021}\natexlab{}.
\newblock \bibinfo{title}{Muppet: Massive Multi-task Representations with
  Pre-Finetuning}.
\newblock
\newblock
\showeprint[arxiv]{2101.11038}~[cs.CL]


\bibitem[{Amazon Web Services, Inc.}(2023a)]%
        {AmazonEC2p4d}
\bibfield{author}{\bibinfo{person}{{Amazon Web Services, Inc.}}}
  \bibinfo{year}{2023}\natexlab{a}.
\newblock \bibinfo{title}{{A}mazon {EC2} {P4d} Instances}.
\newblock
\newblock
\newblock
\shownote{\url{https://aws.amazon.com/ec2/instance-types/p4/}}.


\bibitem[{Amazon Web Services, Inc.}(2023b)]%
        {aws2022elastic}
\bibfield{author}{\bibinfo{person}{{Amazon Web Services, Inc.}}}
  \bibinfo{year}{2023}\natexlab{b}.
\newblock \bibinfo{title}{{Elastic Fabric Adapter}}.
\newblock
\newblock
\newblock
\shownote{\url{https://aws.amazon.com/hpc/efa/}}.


\bibitem[Beltagy et~al\mbox{.}(2020)]%
        {beltagy2020longformer}
\bibfield{author}{\bibinfo{person}{Iz Beltagy}, \bibinfo{person}{Matthew~E.
  Peters}, {and} \bibinfo{person}{Arman Cohan}.}
  \bibinfo{year}{2020}\natexlab{}.
\newblock \showarticletitle{Longformer: The Long-Document Transformer}.
\newblock  (\bibinfo{year}{2020}).
\newblock
\showeprint[arxiv]{2004.05150}~[cs.CL]


\bibitem[Boudoukh et~al\mbox{.}(2001)]%
        {boudoukh2001cyclicschedule}
\bibfield{author}{\bibinfo{person}{Tami Boudoukh}, \bibinfo{person}{Michal
  Penn}, {and} \bibinfo{person}{Gideon Weiss}.}
  \bibinfo{year}{2001}\natexlab{}.
\newblock \showarticletitle{Scheduling jobshops with some identical or similar
  jobs}.
\newblock \bibinfo{journal}{\emph{Journal of Scheduling}} \bibinfo{volume}{4},
  \bibinfo{number}{4} (\bibinfo{year}{2001}), \bibinfo{pages}{177--199}.
\newblock


\bibitem[Brown et~al\mbox{.}(2020)]%
        {brown2020gpt3}
\bibfield{author}{\bibinfo{person}{Tom Brown}, \bibinfo{person}{Benjamin Mann},
  \bibinfo{person}{Nick Ryder}, \bibinfo{person}{Melanie Subbiah},
  \bibinfo{person}{Jared~D Kaplan}, \bibinfo{person}{Prafulla Dhariwal},
  \bibinfo{person}{Arvind Neelakantan}, \bibinfo{person}{Pranav Shyam},
  \bibinfo{person}{Girish Sastry}, \bibinfo{person}{Amanda Askell},
  \bibinfo{person}{Sandhini Agarwal}, \bibinfo{person}{Ariel Herbert-Voss},
  \bibinfo{person}{Gretchen Krueger}, \bibinfo{person}{Tom Henighan},
  \bibinfo{person}{Rewon Child}, \bibinfo{person}{Aditya Ramesh},
  \bibinfo{person}{Daniel Ziegler}, \bibinfo{person}{Jeffrey Wu},
  \bibinfo{person}{Clemens Winter}, \bibinfo{person}{Chris Hesse},
  \bibinfo{person}{Mark Chen}, \bibinfo{person}{Eric Sigler},
  \bibinfo{person}{Mateusz Litwin}, \bibinfo{person}{Scott Gray},
  \bibinfo{person}{Benjamin Chess}, \bibinfo{person}{Jack Clark},
  \bibinfo{person}{Christopher Berner}, \bibinfo{person}{Sam McCandlish},
  \bibinfo{person}{Alec Radford}, \bibinfo{person}{Ilya Sutskever}, {and}
  \bibinfo{person}{Dario Amodei}.} \bibinfo{year}{2020}\natexlab{}.
\newblock \showarticletitle{Language Models are Few-Shot Learners}. In
  \bibinfo{booktitle}{\emph{Proc. of NeurIPS}}, Vol.~\bibinfo{volume}{33}.
  \bibinfo{pages}{1877--1901}.
\newblock


\bibitem[Chen et~al\mbox{.}(2016)]%
        {chen2016training}
\bibfield{author}{\bibinfo{person}{Tianqi Chen}, \bibinfo{person}{Bing Xu},
  \bibinfo{person}{Chiyuan Zhang}, {and} \bibinfo{person}{Carlos Guestrin}.}
  \bibinfo{year}{2016}\natexlab{}.
\newblock \showarticletitle{Training Deep Nets with Sublinear Memory Cost}.
\newblock  (\bibinfo{year}{2016}).
\newblock
\showeprint[arxiv]{1604.06174}~[cs.LG]


\bibitem[Child et~al\mbox{.}(2019)]%
        {child2019generating}
\bibfield{author}{\bibinfo{person}{Rewon Child}, \bibinfo{person}{Scott Gray},
  \bibinfo{person}{Alec Radford}, {and} \bibinfo{person}{Ilya Sutskever}.}
  \bibinfo{year}{2019}\natexlab{}.
\newblock \showarticletitle{Generating Long Sequences with Sparse
  Transformers}.
\newblock  (\bibinfo{year}{2019}).
\newblock
\showeprint[arxiv]{1904.10509}~[cs.LG]


\bibitem[Dao et~al\mbox{.}(2022)]%
        {dao2022flashattention}
\bibfield{author}{\bibinfo{person}{Tri Dao}, \bibinfo{person}{Daniel~Y. Fu},
  \bibinfo{person}{Stefano Ermon}, \bibinfo{person}{Atri Rudra}, {and}
  \bibinfo{person}{Christopher R{\'e}}.} \bibinfo{year}{2022}\natexlab{}.
\newblock \showarticletitle{Flash{A}ttention: Fast and Memory-Efficient Exact
  Attention with {IO}-Awareness}. In \bibinfo{booktitle}{\emph{Proc. of
  NeurIPS}}.
\newblock


\bibitem[Devlin et~al\mbox{.}(2019)]%
        {devlin2019bert}
\bibfield{author}{\bibinfo{person}{Jacob Devlin}, \bibinfo{person}{Ming-Wei
  Chang}, \bibinfo{person}{Kenton Lee}, {and} \bibinfo{person}{Kristina
  Toutanova}.} \bibinfo{year}{2019}\natexlab{}.
\newblock \showarticletitle{{{BERT}}: {{Pre-training}} of {{Deep Bidirectional
  Transformers}} for {{Language Understanding}}}. In
  \bibinfo{booktitle}{\emph{Proc. of {{NAACL}}}}. \bibinfo{publisher}{ACL},
  \bibinfo{pages}{4171--4186}.
\newblock


\bibitem[{Google}(2023)]%
        {GCPa2highgpu1g}
\bibfield{author}{\bibinfo{person}{{Google}}.} \bibinfo{year}{2023}\natexlab{}.
\newblock \bibinfo{title}{{GPU platforms}}.
\newblock
\newblock
\newblock
\shownote{\url{https://cloud.google.com/compute/docs/gpus\#a100-40gb}}.


\bibitem[Gottumukkala et~al\mbox{.}(2020)]%
        {gottumukkala2020dynamic}
\bibfield{author}{\bibinfo{person}{Ananth Gottumukkala},
  \bibinfo{person}{Dheeru Dua}, \bibinfo{person}{Sameer Singh}, {and}
  \bibinfo{person}{Matt Gardner}.} \bibinfo{year}{2020}\natexlab{}.
\newblock \showarticletitle{Dynamic sampling strategies for multi-task reading
  comprehension}. In \bibinfo{booktitle}{\emph{Proc. of ACL}}.
  \bibinfo{publisher}{ACL}, \bibinfo{pages}{920--924}.
\newblock


\bibitem[Graves et~al\mbox{.}(1983)]%
        {graves1983reentrant}
\bibfield{author}{\bibinfo{person}{Stephen~C Graves}, \bibinfo{person}{Harlan~C
  Meal}, \bibinfo{person}{Daniel Stefek}, {and} \bibinfo{person}{Abdel~Hamid
  Zeghmi}.} \bibinfo{year}{1983}\natexlab{}.
\newblock \showarticletitle{Scheduling of Re-Entrant Flow Shops}.
\newblock \bibinfo{journal}{\emph{Journal of operations management}}
  \bibinfo{volume}{3}, \bibinfo{number}{4} (\bibinfo{year}{1983}),
  \bibinfo{pages}{197--207}.
\newblock


\bibitem[Hermann et~al\mbox{.}(2015)]%
        {hermann2015cnndm}
\bibfield{author}{\bibinfo{person}{Karl~Moritz Hermann},
  \bibinfo{person}{Tomás Kociský}, \bibinfo{person}{Edward Grefenstette},
  \bibinfo{person}{Lasse Espeholt}, \bibinfo{person}{Will Kay},
  \bibinfo{person}{Mustafa Suleyman}, {and} \bibinfo{person}{Phil Blunsom}.}
  \bibinfo{year}{2015}\natexlab{}.
\newblock \showarticletitle{Teaching Machines to Read and Comprehend}. In
  \bibinfo{booktitle}{\emph{Proc. of NeurIPS}}. \bibinfo{pages}{1693--1701}.
\newblock


\bibitem[Jacobs et~al\mbox{.}(2023)]%
        {jacobs2023dsulysses}
\bibfield{author}{\bibinfo{person}{Sam~Ade Jacobs}, \bibinfo{person}{Masahiro
  Tanaka}, \bibinfo{person}{Chengming Zhang}, \bibinfo{person}{Minjia Zhang},
  \bibinfo{person}{Shuaiwen~Leon Song}, \bibinfo{person}{Samyam Rajbhandari},
  {and} \bibinfo{person}{Yuxiong He}.} \bibinfo{year}{2023}\natexlab{}.
\newblock \showarticletitle{DeepSpeed Ulysses: System Optimizations for
  Enabling Training of Extreme Long Sequence Transformer Models}.
\newblock  (\bibinfo{year}{2023}).
\newblock
\showeprint[arxiv]{2309.14509}~[cs.LG]


\bibitem[Karmarkar and Karp(1982)]%
        {karmarkar1982differencing}
\bibfield{author}{\bibinfo{person}{Narendra Karmarkar} {and}
  \bibinfo{person}{Richard~M Karp}.} \bibinfo{year}{1982}\natexlab{}.
\newblock \bibinfo{booktitle}{\emph{The differencing method of set
  partitioning}}.
\newblock \bibinfo{publisher}{Computer Science Division (EECS), University of
  California Berkeley}.
\newblock


\bibitem[Karp(2010)]%
        {karp2010reducibility}
\bibfield{author}{\bibinfo{person}{Richard~M Karp}.}
  \bibinfo{year}{2010}\natexlab{}.
\newblock \bibinfo{booktitle}{\emph{Reducibility among combinatorial
  problems}}.
\newblock \bibinfo{publisher}{Springer}.
\newblock


\bibitem[Krell et~al\mbox{.}(2022)]%
        {krell2021packwmask}
\bibfield{author}{\bibinfo{person}{Mario~Michael Krell}, \bibinfo{person}{Matej
  Kosec}, \bibinfo{person}{Sergio~P. Perez}, {and} \bibinfo{person}{Andrew
  Fitzgibbon}.} \bibinfo{year}{2022}\natexlab{}.
\newblock \bibinfo{title}{Efficient Sequence Packing without
  Cross-contamination: Accelerating Large Language Models without Impacting
  Performance}.
\newblock
\newblock
\showeprint[arxiv]{2107.02027}~[cs.CL]


\bibitem[Li et~al\mbox{.}(2023)]%
        {li2023lightseq}
\bibfield{author}{\bibinfo{person}{Dacheng Li}, \bibinfo{person}{Rulin Shao},
  \bibinfo{person}{Anze Xie}, \bibinfo{person}{Eric~P. Xing},
  \bibinfo{person}{Joseph~E. Gonzalez}, \bibinfo{person}{Ion Stoica},
  \bibinfo{person}{Xuezhe Ma}, {and} \bibinfo{person}{Hao Zhang}.}
  \bibinfo{year}{2023}\natexlab{}.
\newblock \showarticletitle{LightSeq: Sequence Level Parallelism for
  Distributed Training of Long Context Transformers}.
\newblock  (\bibinfo{year}{2023}).
\newblock
\showeprint[arxiv]{2310.03294}~[cs.LG]


\bibitem[Longpre et~al\mbox{.}(2023)]%
        {Longpre2022flanv2}
\bibfield{author}{\bibinfo{person}{Shayne Longpre}, \bibinfo{person}{Le Hou},
  \bibinfo{person}{Tu Vu}, \bibinfo{person}{Albert Webson},
  \bibinfo{person}{Hyung~Won Chung}, \bibinfo{person}{Yi Tay},
  \bibinfo{person}{Denny Zhou}, \bibinfo{person}{Quoc~V. Le},
  \bibinfo{person}{Barret Zoph}, \bibinfo{person}{Jason Wei}, {and}
  \bibinfo{person}{Adam Roberts}.} \bibinfo{year}{2023}\natexlab{}.
\newblock \bibinfo{title}{The Flan Collection: Designing Data and Methods for
  Effective Instruction Tuning}.
\newblock
\newblock
\showeprint[arxiv]{2301.13688}~[cs.AI]


\bibitem[Ltd.(2023)]%
        {redis}
\bibfield{author}{\bibinfo{person}{Redis Ltd.}}
  \bibinfo{year}{2023}\natexlab{}.
\newblock \bibinfo{title}{Redis}.
\newblock
\newblock
\urldef\tempurl%
\url{https://redis.io/}
\showURL{%
\tempurl}


\bibitem[Microsoft(2023)]%
        {deepspeed}
\bibfield{author}{\bibinfo{person}{Microsoft}.}
  \bibinfo{year}{2023}\natexlab{}.
\newblock \bibinfo{title}{DeepSpeed}.
\newblock \bibinfo{howpublished}{\url{https://github.com/microsoft/DeepSpeed}}.
\newblock


\bibitem[{Microsoft}(2023)]%
        {MSNDA100}
\bibfield{author}{\bibinfo{person}{{Microsoft}}.}
  \bibinfo{year}{2023}\natexlab{}.
\newblock \bibinfo{title}{{ND A100 v4-series}}.
\newblock
\newblock
\newblock
\shownote{\url{https://learn.microsoft.com/en-us/azure/virtual-machines/nda100-v4-series},
  Last accessed on 2023-10-31}.


\bibitem[Mishra et~al\mbox{.}(2022)]%
        {mishra2022instrtune}
\bibfield{author}{\bibinfo{person}{Swaroop Mishra}, \bibinfo{person}{Daniel
  Khashabi}, \bibinfo{person}{Chitta Baral}, {and} \bibinfo{person}{Hannaneh
  Hajishirzi}.} \bibinfo{year}{2022}\natexlab{}.
\newblock \showarticletitle{Cross-Task Generalization via Natural Language
  Crowdsourcing Instructions}. In \bibinfo{booktitle}{\emph{Proc. of ACL}}.
  \bibinfo{publisher}{ACL}, \bibinfo{pages}{3470--3487}.
\newblock


\bibitem[Narayanan et~al\mbox{.}(2019)]%
        {narayanan2019pipedream}
\bibfield{author}{\bibinfo{person}{Deepak Narayanan}, \bibinfo{person}{Aaron
  Harlap}, \bibinfo{person}{Amar Phanishayee}, \bibinfo{person}{Vivek
  Seshadri}, \bibinfo{person}{Nikhil~R. Devanur}, \bibinfo{person}{Gregory~R.
  Ganger}, \bibinfo{person}{Phillip~B. Gibbons}, {and} \bibinfo{person}{Matei
  Zaharia}.} \bibinfo{year}{2019}\natexlab{}.
\newblock \showarticletitle{{{PipeDream}}: {{Generalized}} Pipeline Parallelism
  for {{DNN}} Training}. In \bibinfo{booktitle}{\emph{Proc. of {{SOSP}}}}.
  \bibinfo{publisher}{ACM}, \bibinfo{pages}{1--15}.
\newblock


\bibitem[Narayanan et~al\mbox{.}(2021)]%
        {narayanan2021efficientmlm}
\bibfield{author}{\bibinfo{person}{Deepak Narayanan}, \bibinfo{person}{Mohammad
  Shoeybi}, \bibinfo{person}{Jared Casper}, \bibinfo{person}{Patrick
  LeGresley}, \bibinfo{person}{Mostofa Patwary}, \bibinfo{person}{Vijay
  Korthikanti}, \bibinfo{person}{Dmitri Vainbrand}, \bibinfo{person}{Prethvi
  Kashinkunti}, \bibinfo{person}{Julie Bernauer}, \bibinfo{person}{Bryan
  Catanzaro}, {et~al\mbox{.}}} \bibinfo{year}{2021}\natexlab{}.
\newblock \showarticletitle{Efficient Large-Scale Language Model Training on
  {{GPU}} Clusters Using {{Megatron-LM}}}. In \bibinfo{booktitle}{\emph{Proc.
  of {{SC}}}}. \bibinfo{publisher}{ACM}, \bibinfo{pages}{1--15}.
\newblock


\bibitem[{NVIDIA}(2023)]%
        {nvidia2023nccl}
\bibfield{author}{\bibinfo{person}{{NVIDIA}}.} \bibinfo{year}{2023}\natexlab{}.
\newblock \bibinfo{title}{{{NCCL}}}.
\newblock
\newblock
\urldef\tempurl%
\url{https://developer.nvidia.com/nccl}
\showURL{%
\tempurl}


\bibitem[Ott et~al\mbox{.}(2019)]%
        {ott2019fairseq}
\bibfield{author}{\bibinfo{person}{Myle Ott}, \bibinfo{person}{Sergey Edunov},
  \bibinfo{person}{Alexei Baevski}, \bibinfo{person}{Angela Fan},
  \bibinfo{person}{Sam Gross}, \bibinfo{person}{Nathan Ng},
  \bibinfo{person}{David Grangier}, {and} \bibinfo{person}{Michael Auli}.}
  \bibinfo{year}{2019}\natexlab{}.
\newblock \showarticletitle{Fairseq: {{A}} Fast, Extensible Toolkit for
  Sequence Modeling}. In \bibinfo{booktitle}{\emph{Proc. of {{NAACL}}}}.
  \bibinfo{publisher}{ACL}.
\newblock


\bibitem[Paszke et~al\mbox{.}(2019)]%
        {paszke2019pytorch}
\bibfield{author}{\bibinfo{person}{Adam Paszke}, \bibinfo{person}{Sam Gross},
  \bibinfo{person}{Francisco Massa}, \bibinfo{person}{Adam Lerer},
  \bibinfo{person}{James Bradbury}, \bibinfo{person}{Gregory Chanan},
  \bibinfo{person}{Trevor Killeen}, \bibinfo{person}{Zeming Lin},
  \bibinfo{person}{Natalia Gimelshein}, \bibinfo{person}{Luca Antiga},
  \bibinfo{person}{Alban Desmaison}, \bibinfo{person}{Andreas Kopf},
  \bibinfo{person}{Edward Yang}, \bibinfo{person}{Zachary DeVito},
  \bibinfo{person}{Martin Raison}, \bibinfo{person}{Alykhan Tejani},
  \bibinfo{person}{Sasank Chilamkurthy}, \bibinfo{person}{Benoit Steiner},
  \bibinfo{person}{Lu Fang}, \bibinfo{person}{Junjie Bai}, {and}
  \bibinfo{person}{Soumith Chintala}.} \bibinfo{year}{2019}\natexlab{}.
\newblock \showarticletitle{{{PyTorch}}: {{An Imperative Style}},
  {{High-Performance Deep Learning Library}}}. In
  \bibinfo{booktitle}{\emph{Proc. of {{NeurIPS}}}}.
  \bibinfo{pages}{8024--8035}.
\newblock


\bibitem[Raffel et~al\mbox{.}(2020)]%
        {raffel2020t5}
\bibfield{author}{\bibinfo{person}{Colin Raffel}, \bibinfo{person}{Noam
  Shazeer}, \bibinfo{person}{Adam Roberts}, \bibinfo{person}{Katherine Lee},
  \bibinfo{person}{Sharan Narang}, \bibinfo{person}{Michael Matena},
  \bibinfo{person}{Yanqi Zhou}, \bibinfo{person}{Wei Li},
  \bibinfo{person}{Peter~J Liu}, {et~al\mbox{.}}}
  \bibinfo{year}{2020}\natexlab{}.
\newblock \showarticletitle{Exploring the limits of transfer learning with a
  unified text-to-text transformer.}
\newblock \bibinfo{journal}{\emph{Journal of Machine Learning Research}}
  \bibinfo{volume}{21}, \bibinfo{number}{140} (\bibinfo{year}{2020}),
  \bibinfo{pages}{1--67}.
\newblock


\bibitem[Rajbhandari et~al\mbox{.}(2020)]%
        {rajbhandari2020zero}
\bibfield{author}{\bibinfo{person}{Samyam Rajbhandari}, \bibinfo{person}{Jeff
  Rasley}, \bibinfo{person}{Olatunji Ruwase}, {and} \bibinfo{person}{Yuxiong
  He}.} \bibinfo{year}{2020}\natexlab{}.
\newblock \showarticletitle{{{ZeRO}}: Memory optimizations toward training
  trillion parameter models}. In \bibinfo{booktitle}{\emph{Proc. of SC}}. IEEE,
  \bibinfo{pages}{1--16}.
\newblock


\bibitem[Sanh et~al\mbox{.}(2022)]%
        {sanh2022t0}
\bibfield{author}{\bibinfo{person}{Victor Sanh}, \bibinfo{person}{Albert
  Webson}, \bibinfo{person}{Colin Raffel}, \bibinfo{person}{Stephen Bach},
  \bibinfo{person}{Lintang Sutawika}, \bibinfo{person}{Zaid Alyafeai},
  \bibinfo{person}{Antoine Chaffin}, \bibinfo{person}{Arnaud Stiegler},
  \bibinfo{person}{Teven Le~Scao}, \bibinfo{person}{Arun Raja},
  {et~al\mbox{.}}} \bibinfo{year}{2022}\natexlab{}.
\newblock \showarticletitle{Multitask Prompted Training Enables Zero-Shot Task
  Generalization}. In \bibinfo{booktitle}{\emph{Proc. of ICLR}}.
  \bibinfo{publisher}{OpenReview.net}.
\newblock


\bibitem[Vaswani et~al\mbox{.}(2018)]%
        {tensor2tensor}
\bibfield{author}{\bibinfo{person}{Ashish Vaswani}, \bibinfo{person}{Samy
  Bengio}, \bibinfo{person}{Eugene Brevdo}, \bibinfo{person}{Francois Chollet},
  \bibinfo{person}{Aidan~N. Gomez}, \bibinfo{person}{Stephan Gouws},
  \bibinfo{person}{Llion Jones}, \bibinfo{person}{Łukasz Kaiser},
  \bibinfo{person}{Nal Kalchbrenner}, \bibinfo{person}{Niki Parmar},
  \bibinfo{person}{Ryan Sepassi}, \bibinfo{person}{Noam Shazeer}, {and}
  \bibinfo{person}{Jakob Uszkoreit}.} \bibinfo{year}{2018}\natexlab{}.
\newblock \showarticletitle{Tensor2Tensor for Neural Machine Translation}.
\newblock  (\bibinfo{year}{2018}).
\newblock
\showeprint[arxiv]{1803.07416}~[cs.LG]


\bibitem[Vaswani et~al\mbox{.}(2017)]%
        {vaswani2017attention}
\bibfield{author}{\bibinfo{person}{Ashish Vaswani}, \bibinfo{person}{Noam
  Shazeer}, \bibinfo{person}{Niki Parmar}, \bibinfo{person}{Jakob Uszkoreit},
  \bibinfo{person}{Llion Jones}, \bibinfo{person}{Aidan~N Gomez},
  \bibinfo{person}{{\L}ukasz Kaiser}, {and} \bibinfo{person}{Illia
  Polosukhin}.} \bibinfo{year}{2017}\natexlab{}.
\newblock \showarticletitle{Attention is all you need}.
\newblock \bibinfo{journal}{\emph{Proc. of NeurIPS}}  \bibinfo{volume}{30}
  (\bibinfo{year}{2017}).
\newblock


\bibitem[Wei et~al\mbox{.}(2022)]%
        {wei2022flan}
\bibfield{author}{\bibinfo{person}{Jason Wei}, \bibinfo{person}{Maarten Bosma},
  \bibinfo{person}{Vincent Zhao}, \bibinfo{person}{Kelvin Guu},
  \bibinfo{person}{Adams~Wei Yu}, \bibinfo{person}{Brian Lester},
  \bibinfo{person}{Nan Du}, \bibinfo{person}{Andrew~M. Dai}, {and}
  \bibinfo{person}{Quoc~V Le}.} \bibinfo{year}{2022}\natexlab{}.
\newblock \showarticletitle{Finetuned Language Models are Zero-Shot Learners}.
  In \bibinfo{booktitle}{\emph{Proc. of ICLR}}.
  \bibinfo{publisher}{OpenReview.net}.
\newblock


\bibitem[Williams et~al\mbox{.}(2018)]%
        {williams2018mnli}
\bibfield{author}{\bibinfo{person}{Adina Williams}, \bibinfo{person}{Nikita
  Nangia}, {and} \bibinfo{person}{Samuel Bowman}.}
  \bibinfo{year}{2018}\natexlab{}.
\newblock \showarticletitle{A Broad-Coverage Challenge Corpus for Sentence
  Understanding through Inference}. In \bibinfo{booktitle}{\emph{Proc. of
  {{NAACL}}}}. \bibinfo{publisher}{ACL}, \bibinfo{pages}{1112--1122}.
\newblock


\bibitem[Zhai et~al\mbox{.}(2023)]%
        {zhai2023bytetransformer}
\bibfield{author}{\bibinfo{person}{Yujia Zhai}, \bibinfo{person}{Chengquan
  Jiang}, \bibinfo{person}{Leyuan Wang}, \bibinfo{person}{Xiaoying Jia},
  \bibinfo{person}{Shang Zhang}, \bibinfo{person}{Zizhong Chen},
  \bibinfo{person}{Xin Liu}, {and} \bibinfo{person}{Yibo Zhu}.}
  \bibinfo{year}{2023}\natexlab{}.
\newblock \showarticletitle{{ByteTransformer}: A high-performance transformer
  boosted for variable-length inputs}. In \bibinfo{booktitle}{\emph{Proc. of
  {{IPDPS}}}}. \bibinfo{publisher}{IEEE}, \bibinfo{pages}{344--355}.
\newblock


\bibitem[Zhang et~al\mbox{.}(2022)]%
        {zhang2022mics}
\bibfield{author}{\bibinfo{person}{Zhen Zhang}, \bibinfo{person}{Shuai Zheng},
  \bibinfo{person}{Yida Wang}, \bibinfo{person}{Justin Chiu},
  \bibinfo{person}{George Karypis}, \bibinfo{person}{Trishul Chilimbi},
  \bibinfo{person}{Mu Li}, {and} \bibinfo{person}{Xin Jin}.}
  \bibinfo{year}{2022}\natexlab{}.
\newblock \showarticletitle{MiCS: Near-Linear Scaling for Training Gigantic
  Model on Public Cloud}.
\newblock \bibinfo{journal}{\emph{Proc. VLDB Endow.}} \bibinfo{volume}{16},
  \bibinfo{number}{1} (\bibinfo{year}{2022}), \bibinfo{pages}{37–50}.
\newblock


\bibitem[Zhao et~al\mbox{.}(2023)]%
        {zhao2023fsdp}
\bibfield{author}{\bibinfo{person}{Yanli Zhao}, \bibinfo{person}{Andrew Gu},
  \bibinfo{person}{Rohan Varma}, \bibinfo{person}{Liang Luo},
  \bibinfo{person}{Chien-Chin Huang}, \bibinfo{person}{Min Xu},
  \bibinfo{person}{Less Wright}, \bibinfo{person}{Hamid Shojanazeri},
  \bibinfo{person}{Myle Ott}, \bibinfo{person}{Sam Shleifer},
  \bibinfo{person}{Alban Desmaison}, \bibinfo{person}{Can Balioglu},
  \bibinfo{person}{Pritam Damania}, \bibinfo{person}{Bernard Nguyen},
  \bibinfo{person}{Geeta Chauhan}, \bibinfo{person}{Yuchen Hao},
  \bibinfo{person}{Ajit Mathews}, {and} \bibinfo{person}{Shen Li}.}
  \bibinfo{year}{2023}\natexlab{}.
\newblock \showarticletitle{PyTorch FSDP: Experiences on Scaling Fully Sharded
  Data Parallel}.
\newblock  (\bibinfo{year}{2023}).
\newblock
\showeprint[arxiv]{2304.11277}~[cs.DC]


\bibitem[Zheng et~al\mbox{.}(2022)]%
        {zheng2022alpa}
\bibfield{author}{\bibinfo{person}{Lianmin Zheng}, \bibinfo{person}{Zhuohan
  Li}, \bibinfo{person}{Hao Zhang}, \bibinfo{person}{Yonghao Zhuang},
  \bibinfo{person}{Zhifeng Chen}, \bibinfo{person}{Yanping Huang},
  \bibinfo{person}{Yida Wang}, \bibinfo{person}{Yuanzhong Xu},
  \bibinfo{person}{Danyang Zhuo}, \bibinfo{person}{Eric~P Xing},
  {et~al\mbox{.}}} \bibinfo{year}{2022}\natexlab{}.
\newblock \showarticletitle{Alpa: Automating inter- and intra-operator
  parallelism for distributed deep learning}. In
  \bibinfo{booktitle}{\emph{Proc. of OSDI}}. \bibinfo{publisher}{USENIX},
  \bibinfo{pages}{559--578}.
\newblock


\end{thebibliography}
